 \documentclass[acmlarge,screen]{acmart}

\usepackage{array}
\usepackage{multirow}
\usepackage{lscape}
\usepackage{float}
\usepackage{tabularx}
\usepackage{makecell}
\usepackage{graphicx}
\usepackage{color}
\usepackage{booktabs}
\usepackage{ragged2e}
\usepackage{ulem}
\AtBeginDocument{%
  \providecommand\BibTeX{{%
    \normalfont B\kern-0.5em{\scshape i\kern-0.25em b}\kern-0.8em\TeX}}}

\newcommand{\inlinequote}[1]{{{\it ``#1''}}}

\newcommand{\jtrevised}[1]{\textcolor{black}{#1}}
\newcommand{\kxrevised}[1]{\textcolor{black}{#1}}
\newcommand{\joeyrevised}[1]{\textcolor{black}{#1}}
\newcommand{\hkrevised}[1]{\textcolor{black}{#1}}

\setcopyright{acmcopyright}
\copyrightyear{2024}
\acmYear{2024}
\acmDOI{XXXXXXX.XXXXXXX}

\acmConference[Conference acronym 'XX]{Make sure to enter the correct
  conference title from your rights confirmation emai}{June 03--05,
  2024}{Woodstock, NY}
\acmBooktitle{Woodstock '24: ACM Symposium on Neural Gaze Detection,
 June 03--05, 2024, Woodstock, NY} 
\acmPrice{15.00}
\acmISBN{978-1-4503-XXXX-X/18/06}

\begin{document}

%%
%% The "title" command has an optional parameter,
%% allowing the author to define a "short title" to be used in page headers.
\title[Understanding Current Practices of AT's Customized Modification in China]{"I see it as a wellspring for my positive and upward journey in life.": Understanding Current Practices of Assistive Technology's Customized Modification in China}

%%
%% The "author" command and its associated commands are used to define
%% the authors and their affiliations.
%% Of note is the shared affiliation of the first two authors, and the
%% "authornote" and "authornotemark" commands
%% used to denote shared contribution to the research.
\author{Kexin Yang}
\authornote{The authors contributed equally to this research.}
%\email{trovato@corporation.com}
%\orcid{1234-5678-9012}
\author{Junyi Wu}
\authornotemark[1]
\author{Haokun Xin}
\authornotemark[1]
%\email{webmaster@marysville-ohio.com}
\affiliation{%
  \institution{Institute for AI Industry Research, Tsinghua University}
  \city{Beijing}
  \country{China}
}

\author{Jiangtao Gong}
\authornote{Corresponding Author}
\email{gongjiangtao2@gmail.com}
\orcid{0000-0002-4310-1894}
\affiliation{%
  \institution{Institute for AI Industry Research, Tsinghua University}
  \city{Beijing}
  \country{China}
}

%%
%% By default, the full list of authors will be used in the page
%% headers. Often, this list is too long, and will overlap
%% other information printed in the page headers. This command allows
%% the author to define a more concise list
%% of authors' names for this purpose.
\renewcommand{\shortauthors}{Yang, Wu, and Xin, et al.}

%%
%% The abstract is a short summary of the work to be presented in the
%% article.
\begin{abstract}
%因为残障人士的身体情况和生活环境差异大，标准化辅具经常无法满足他们的需求。DIY辅具在很多高收入国家是一个流行的解决方案，然而对low-income，特别是慈善文化不发达的中国，尚缺乏document。在这篇论文中，我们深入地访谈了XX个使用DIY辅具的残疾人和XX个提供DIY辅具的相关人士来了解这一现状。基于调研的结果，我们梳理了目前中国残疾人DIY辅具的一般流程、DIY辅具带来的好处。我们发现DIY辅具不止是让残疾人身体更舒适，生活更便捷，它还为残疾人带来了自信，减轻了社会压力，甚至帮助残疾人完成了自我实现。此外，我们还总结了残疾人在辅具改装前、中、后遇到的挑战。特别地，我们分析报告了目前中国DIY辅具的典型的商业模式及困境。我们的研究对于未来AT的普惠化个性化生产提供了重要的设计依据和研究insight。
  Due to the significant differences in physical conditions and living environments of people with disabilities, standardized assistive technologies (ATs) often fail to meet their needs. Modified AT, especially DIY (Do It Yourself) ATs, are a popular solution in many high-income countries, but there is a lack of documentation for low- and middle-income areas, especially in China, where the culture of philanthropy is undeveloped. To understand the current situation in this paper, we conducted semi-structured interviews with 10 individuals with disabilities using modified ATs and 10 individuals involved in providing these including family members, standard assistive device manufacturers, and individuals employed for their modification skills, etc. Based on the results of the thematic analysis, we have summarized the general process of modified ATs for people with disabilities in China and the benefits these devices bring. We found that modified ATs not only make the lives of people with disabilities more comfortable and convenient but also bring them confidence, reduce social pressure, and even help them achieve self-realization. 
  %~\kxrevised{Also, there are three main challenges for preparing for a modification: awareness gap, family resistance, and organizational limits. People with disabilities physical conditions limit modifications, difficulty in finding the right assistant, and limited resources and craftsmanship are three operation challenges during the modification. Furthermore, there were three main experiential challenges in the use phase after completing the modification: physiological discomfort in use, more trouble for family members with the modified aids, and difficulty in adapting the aids to complex external environments.} 
  ~\jtrevised{Additionally, we summarized the challenges they encountered before, during, and after the modification, including awareness gaps, family resistance, a lack of a business model, and so on.}
  Specifically, we conducted a special case study about the typical business models and challenges currently faced by AT Modification Organizations in China. Our research provides important design foundations and research insights for the future of universal and personalized production of AT.
\end{abstract}

%%
%% The code below is generated by the tool at http://dl.acm.org/ccs.cfm.
%% Please copy and paste the code instead of the example below.
%%

%%
%% Keywords. The author(s) should pick words that accurately describe
%% the work being presented. Separate the keywords with commas.

%% A "teaser" image appears between the author and affiliation
%% information and the body of the document, and typically spans the
%% page.
\begin{teaserfigure}
  \includegraphics[width=1\textwidth]{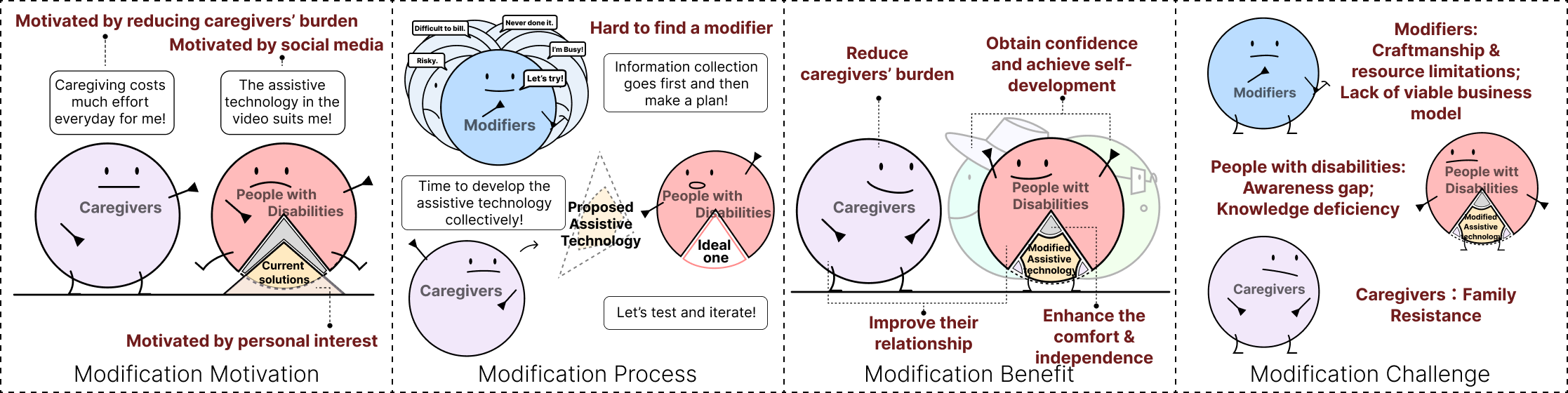}
  \caption{Current Practices of Assistive Technology's Customized Modification in China.}
  \Description{Enjoying the baseball game from the third-base
  seats. Ichiro Suzuki preparing to bat.}
  \label{fig:teaser}
\end{teaserfigure}

% 04/24/24 Figure 1 is cute, yet The reviewer don’t understand the meaning of the “no no no no yes” figure in the modification process. Another important aspect that should be included in this figure is the modification risk if the empirical data shows any. The author can include this in the “modification challenges” section. It's a very smart idea to show this figure at the beginning of the paper.要加个数字进来，nonono看不太懂

% \received{20 February 2007}
% \received[revised]{12 March 2009}
% \received[accepted]{5 June 2009}

%%
%% This command processes the author and affiliation and title
%% information and builds the first part of the formatted document.
\maketitle

\section{Introduction}%（第一段background，讲这件事情为什么重要，第二段现在的gap）-三段稍微多点，research questions，第三段是方法和findings，总结我们的contribution，our contribution can summarized as 1, 2, 3
Assistive Technology (AT) plays a crucial role in enabling individuals to perform tasks that would otherwise be impossible or more difficult~\cite{mccreadie_tinker_2005}, thereby enhancing their independence and autonomy~\cite{Zupan2012Assistive,zhang2023follower}, improving their functionality across various life settings~\cite{Johnson1997Assistive}, and overall quality of life~\cite{8862454}. However, the specialized nature of AT leads to high development costs, limited market demand, and consequently, steep retail prices ~\cite{de2011design}. Furthermore, the diverse and evolving needs of individuals with disabilities often render generalized AT solutions inadequate~\cite{kintsch2002framework}. Modified AT adapts existing ATs to better fit individual needs, often including Do-It-Yourself Assistive Technology (DIY-AT) and customized-AT. Customized-AT, due to its professional requirements, differs from the more accessible and affordable DIY-AT.In regions where research on the applicability of AT is more extensive, there has been a significant emergence of research on a typical way of customized modification, DIY-AT, which refers to the creation and adaptation of AT by non-professionals, including people with disabilities\kxrevised{~\cite{hook2014study,hurst2011empowering,meissner2017yourself,gong2020helicoach,ding2019helicoach}}. Coupled with the integration of new technologies, this has led to the establishment of robust, socially supported online platforms and systematic services for DIY-AT~\cite{unknown2019development}. Understanding that modified AT is a way with abundant potential benefits, many resources, services, and technologies have been developed to help people.

An increasing body of research within the domains of Human-Computer Interaction and Computer-Supported Cooperative Work (CSCW) is giving significant attention to the challenges and solutions in designing ATs collaboratively~\cite{chang2022assistive,baldwin2019design,parry2017understanding,slegers2020makers}, especially the rise of DIY-ATs in low- and middle-income countries.~\cite{hamidi2022knowledge, 10.1145/3359257}.~\jtrevised{DIY-ATs enable individuals to engage socially without constantly depending on others for assistance~\cite{morte2020personal}. With a sustainable business model, providing tools and devices tailored to their needs can facilitate communication, access to information, and participation in social activities~\cite{pedersen2021fact}. Due to the positive impact that AT has on communication and collaboration with others, in-depth research on AT is essential. However, even in economically developed areas (countries with a high Human
Development Index (HDI) such as the United States, Canada, and Denmark), the business models for AT are equally challenging due to the small market size and high costs of customization~\cite{oderanti2016holistic,Kumar2023SolutionFocusedAT,borg2011assistive}.} 

In China, the current state of DIY-AT remains unclear. 
Especially due to the influence of traditional attitudes towards disabilities, individuals with disabilities encounter certain difficulties in integrating into society both psychologically and in their daily mobility in china~\cite{campbell2011invisibles,bruyere2018disability}. It is challenging for people to encounter individuals with disabilities in public settings such as workplaces and schools~\footnote{https://www.sixthtone.com/news/1001285}. This backdrop has led to relatively limited and superficial research and technology services for people with disabilities in China~\cite{li2021choose}. In addition, individuals in low- and middle-income countries have limited access to the most appropriate AT and often lack a comprehensive understanding of its capabilities~\cite{eide2009assistive,may1999survey}. DIY-AT serves as a solution to the current challenges faced by people with disability, especially in these countries. According to a survey conducted by the China Disabled Persons' Federation (CDPF) in 2021, over one-third of the country's 82 million individuals with disabilities were identified as having a significant need for suitable ATs~\footnote{https://www.cdpf.org.cn/zwgk/zccx/cjrgk/93a052e1b3d342ed8a059357cabf09ca.htm}. There is a significant demand and market for AT that meet specific individual requirements and are suitable for personal use. % 04/24/24 因为人口多……这个句子需要一个参考。后来的作者提出了“AT的需求和市场很大”的说法，这可能不是真的。仅仅因为有大量的群体需要这些技术并不意味着它有很高的市场需求。确实需要很高，但人们对这些技术的消费态度可能非常低，这些技术通常很昂贵，或者这就是为什么人们“占用现有技术”，他们可以负担得起并拥有，甚至像把一些布包裹在手杖的手柄上，例如。 人口多不等于市场大。 寻找能够支持中国个性化辅具市场很大的论文。
 This situation underscores the importance of developing and providing a wide range of ATs that can be adapted or customized to address the diverse needs of this population. Nevertheless, the impact and challenges faced by individuals with disabilities in China in the context of DIY-AT are not well understood due to lack of research and the economic situation of persons with disabilities in China. Our research aims to address the following research questions: \textbf{1. What are the motivations, specific processes, benefits, and challenges of individuals with disabilities in China engaging in DIY-AT? 2. In the context of China, how do the existing business models for DIY-AT and social networks around people with disabilities for modified ATs operate?}

To address our research questions, we conducted semi-structured interviews with 20 participants, comprising four distinct groups: individuals with disabilities who use and are involved in the modification of ATs; individuals with experience in modifying ATs for others; representatives from third-party modification agencies and non-profit organizations; and personnel from the accessibility departments of the China Disabled Persons’ Federations (CDPF). 
%From this study, we discuss the motivations for DIY-AT among individuals with disabilities as influenced by ~\kxrevised{social media in China. We then discuss the psychological impacts and influence on solitude and socialization that the practice of DIY-AT brings to these individuals. We highlight the challenges faced by people with disabilities during the modification process.}
\jtrevised{
We meticulously analyzed the modification process, the benefits it brings, and the challenges faced by people with disabilities during the modification process. Specifically, we conducted a case study on the current business model for AT modifications for people with disabilities in China. Based on these findings, we discuss the importance of AT modifications for people with disabilities in China, the business model, and have proposed design implications.}

%In the Chinese context, AT modification can partly address the unmet personalized needs of people with disabilities. The exact application is still in doubt. 
Thus, our research contributes in the following ways: i) Detailed motivations and benefits of AT modifications among Chinese people with disabilities; ii) Systematically summarizing of the current process of AT modification in China with adapted the Double Diamond design model framework\cite{council2005double}; iii) Analysis of the challenges faced by persons with disabilities in China during the before, during, and after modification; iv) A specialized case study on the current business models for AT modification in China; v) Based on our findings, proposed suggestions for future AT modifications. Our research offers important insights for government policies to incentivize the AT modification market and for the scaling of AT modifications.
%i) Breaking Down the AT Modification Process: We systematically analyze how individuals with disabilities modify assistive technologies to meet their specific needs; ii) Examining Multi-Stakeholder Dynamics in China's DIY-AT: We explore the interaction and balance among various stakeholders involved in the DIY-AT process, including users, caregivers, and technology suppliers; iii) Benefits bring Enhancing User Participation and Governmental Support: We propose design considerations to increase the involvement of people with disabilities in the AT design process and discuss potential government measures to better meet their needs.

%backgound,这件事重要
%讲gap->RQ
%方法和大概finding
%contribution

\section{Related Work}
%RW写作目的：咱们读了很多论文（80篇）咱们的这个研究很有意义
\subsection{DIY and Modification of AT} 
The needs of people with disabilities for ATs are specific, complex, and changing.  Past researches indicate that standard ATs often fail to meet the specific needs of people with disabilities~\cite{de2018assistive,dawe2006desperately,kintsch2002framework}. For instance, cognitive abilities can differ greatly among individuals with the same disability~\cite{dawe2006desperately}. This diversity in abilities necessitates that ATs be specifically tailored to each individual's unique needs, ensuring these tools effectively enhance their abilities~\cite{kintsch2002framework}. Individuals with disabilities frequently receive ATs that only approximately meet their needs~\cite{dawe2006desperately,de2011design,elmannai2017sensor}.

In the context of generalized AT, several key factors contribute to its under-utilization and abandonment, such as various life situations~\cite{copley2004barriers,boot2018access,phillips1993predictors,handbookassistive}, social acceptability~\cite{hocking1999function}, the high cost of commercial AT~\cite{jaeger2005understanding, carey2004assistive}. %Firstly, commercially available AT often caters to the fixed needs and general circumstances of the majority, disregarding the unique and fluctuating needs of individual users in various life situations~\cite{copley2004barriers,boot2018access,phillips1993predictors,handbookassistive}. This leads to the physical impracticality of using such devices without modifications~\cite{kintsch2002framework,ossmann2012asterics}. From a psychological perspective, the uniform appearance of ATs raises concerns about social acceptability, significantly impacting the user's willingness to use the AT~\cite{hocking1999function}. Economically, the limited AT market contributes to the high cost of AT devices~\cite{jaeger2005understanding}. Most people with disabilities depend on scarce financial support for AT~\cite{carey2004assistive}. This financial strain presents a considerable obstacle for people with disabilities and their caregivers~\cite{ossmann2012asterics}. Consequently, consumers often resort to using less expensive products that fail to meet their needs~\cite{homereducing}. 
These issues often lead to significant consequences, particularly a high rate of AT abandonment~\cite{federici2016abandonment,sugawara2018abandonment}. On an individual level, non-use of a device can lead to decreased functional abilities, reduced freedom and independence, and increased financial burdens~\cite{hocking1999function}. On a service delivery level, device abandonment signifies the ineffective allocation of limited resources by federal, and local government agencies, and other provider organizations~\cite{andrew1990toward}.

Previous research has argued that the development of technologies and services that enable people to design, make, and adapt their own DIY-AT has the potential to address these challenges~\cite{de2011role,hurst2013making}. In numerous instances, the economic advantage of DIY-AT becomes evident, allowing individuals to acquire the necessary equipment with significantly lower financial investment\cite{de2011design,hurst2011empowering}. On the other hand, DIY-AT allows customization, providing potential solutions when off-the-shelf products are inadequate. 

However, the DIY-AT approach has also shown many issues in past research. Due to the lack of technical expertise and experience, DIY efforts can consume a considerable amount of time without necessarily yielding useful devices~\cite{hook2014study}. Additionally, the safety and reliability of DIY-AT are often compromised due to the absence of professional involvement~\cite{hook2014study}. This highlights the critical need for more structured guidance and support in the DIY-AT process to ensure that the end products are not only innovative but also safe and functional for the end-users.

%Motivation为什么要做？
%以往研究中做了哪些DIY?有什么好处和问题？
%research gap:都是欧美国家，技术门槛/成本比较高，对于中国这样的发展中国家，残疾人比较贫困，需要在本文中进行调研。

\subsection{Business Model of AT}
%发达国家的公益组织的运作model，发展中国家的运作model，特别到中国的运作model/志愿者（文化背景），捐赠人，法律政策。
%research gap：中国没有成熟的公益社会援助的文化传统和政治制度，辅具的DIY业务如何商业化运作，有什么挑战，目前来说并不清楚，需要在本文中进行调研。
\hkrevised{A viable business model is crucial for sustainable AT services. In the UK, despite significant investment in the health and social care sector, establishing a sustainable business model remains a significant challenge~\cite{oderanti2016holistic}. In low and middle-income countries (LMICs), AT services typically receive less government funding and charitable donations~\cite{Kumar2023SolutionFocusedAT}. Additionally, the lack of trained professionals to implement AT service education and maintenance~\cite{smith2023overview}, legal standards requiring minimum quality for assistive devices~\cite{pearlman2008lower}, a systematic AT distribution service system~\cite{andrich2013service}, and widespread awareness of assistive technology~\cite{tangcharoensathien2018improving} are factors that further limit the development of AT services in these regions. This is also reflected in the lower prevalence of assistive devices. According to WHO data, countries with a high HDI such as the United States, Canada, and Denmark have an AT acquisition rate as high as 87.7\%\footnote{https://www.who.int/publications/i/item/9789240049178}, while in LMICs, only 5-15\% of people with disabilities can access ATs~\cite{borg2011assistive}.}

\hkrevised{For individuals with disabilities, there are typically three pathways to access AT. The first is purchasing privately or from others at their own expense, the second is receiving government subsidies, and the third is obtaining assistance from charitable organizations~\cite{karki2024processes}. For individuals with disabilities who purchase assistive devices at their own expense, the high cost of these devices is the first challenge. Even in developed countries like the United States, many assistive devices are expensive, and without full or partial subsidies from third-party sources, many people still cannot afford them~\cite{tangcharoensathien2018improving}. In LMICs, individuals with disabilities find it even harder to afford the high cost of assistive devices~\cite{Kumar2023SolutionFocusedAT}. Additionally, policy funding and charitable donations are usually scarcer in LMICs, making it difficult for governments and charitable organizations to provide AT on a large scale. Furthermore, the culture of charity is also less developed in LMICs~\cite{Kumar2023SolutionFocusedAT,smith2023overview}. }

\hkrevised{Typically, providing assistive technology (AT) for individuals with disabilities is considered a governmental responsibility. However, in many cases, especially in remote areas, due to difficulties in securing government funding, individuals with disabilities often opt to pay out of pocket or seek help from charitable organizations~\cite{tam2003survey, karki2024processes}. For charitable organizations, the culture of charity has a significant impact on sources of income.} The development of NGOs in Western contexts is more mature, organized, and professional~\cite{putnam2000bowling}, with a strong cultural atmosphere of charity and altruism, such as in American society, where philanthropy is a longstanding tradition~\cite{putnam2000bowling}. In the United States and other developed countries, charitable activities have become integrated into everyday life. Requests for time or monetary donations are not considered "dirty words"~\cite{kapoor2010fdi}. A study on American NGOs indicates that charitable donations and fundraising are their primary sources of income~\cite{carroll2009revenue}. However, in emerging economies, the infrastructure for philanthropy is still underdeveloped, including in countries like Brazil, Russia, India, China, and South Africa (BRICS), as well as Mexico, Indonesia, and others~\cite{wong2020american}. Taking China as an example, in 2015, the Charities Aid Foundation Global Giving Index ranked China 144th out of 145 countries in terms of donations to strangers, time, and money~\cite{bies2019state}. 

Research indicates that in 2018, there were 2,562 registered NGOs in China dedicated to disability-related initiatives~\cite{wu2022mapping}. NGOs focusing on modified ATs are among these organizations. In recent years, despite the enactment of new laws and regulations such as the "Disability Protection Law," which provides legal support for the expansion of assistive technology services to people with disabilities in China\cite{li2019improving}, and the "Charity Law," which is a sign of the rapid maturation of China's charitable laws and regulations and has eased the burden of NGO dual registration and established mechanisms for private foundations to raise funds from the public\cite{bies2019state}, NGOs dedicated to the cause of people with disabilities in China continue to develop further. However, funding remains a significant challenge for many NGOs~\cite{wang2001development,weller2004civil}, particularly those that are not government-led. Therefore, there is a greater need for viable business models for Chinese NGOs dedicated to modified ATs. These models are essential to sustain the public services they provide. However, there has been a lack of research on the specific business models employed by Chinese NGOs in the past. Therefore, this paper aims to investigate and address this gap in our understanding of these NGOs' business models.

\subsection{AT Modification Needs in developing countries}
Past research has included many open-sharing communities that have shared skills to address potential problems and further encourage DIY-AT~\cite{de2011design,buehler2015sharing,quintero2022review, unknown}. For instance, Thingiverse.com is an online community that supports personalized construction~\cite{buehler2015sharing}. A significant number of these designs are created by end-users themselves or by their friends and loved ones, gaining additional insights through communication with designers and rehabilitation professionals~\cite{buehler2015sharing}. Rapid prototyping, as a more tailored approach, is widely used in the United States~\cite{tian2017making,hofmann2015making,volonghi20183d}. However, these emerging communities and technological advancements that could further aid DIY-AT are predominantly found in developed countries~\cite{buehler2015sharing,hamidi2014yourself,hofmann2019occupational}. In LMICs, the ability to produce assistive technologies locally may be limited due to the lack of technological innovation and expertise, and the high cost of imported products~\cite{tangcharoensathien2018improving}. Importantly, these technologies are developed without taking into account the real environmental, and social factors that impede the adoption of technologies in resource-poor environments~\footnote{https://www.engineeringforchange.org/news/assistive-devices-in-low-income-countries-adaptability-and-recent-innovations/}.

Eighty percent of people with disabilities live in the developing world, and the World Health Organization estimates that only about one in ten individuals in need of AT can access it~\cite{officer2011world}. In the face of such immense demand, developing countries face numerous challenges in AT, including limited policy management for supply and provision~\cite{mcsweeney2019wheelchair}, limited training capabilities, limited product supply and implementation~\cite{banks2017poverty}. The study in Afghanistan and Uganda revealed the current situation of difficulty in accessing appropriate ATs for persons with disabilities and the lack of economic resources are the biggest factors hindering access to ATs~\cite{franccois1998causes,may1999survey}. Therefore, for a large portion of the developing world, providing suitable products for people who need AT is a significant challenge.
%Chinese people with disabilities face multiple challenges, such as barriers in healthcare~\cite{kang2016health}. China's vast wealth gap, particularly between rural and urban people with disabilities, is stark, with a higher poverty rate and less access to medical resources and stable infrastructure in rural areas~\cite{guo2020inequality}. Non-profit organizations, constrained by fundraising, are unable to consistently meet the needs of people with disabilities~\footnote{\url{https://www.cecc.gov/publications/commission-analysis/ongoing-challenges-faced-by-persons-with-disabilities-in-the-people#_ftn14}}. Under these circumstances, it is very difficult for people with disabilities in China to access appropriate ATs. DIY, especially in developing and lower-middle-income countries like China, has a strong demand and a vast market due to its cost-effective nature. 
However, there is currently a lack of empirical research in this area. From 1995 to 2017, only 52 articles were published on knowledge about assistive devices in low- and middle-income countries, and there was little mention of assistive devices other than prosthetics and manual wheelchairs in these articles~\cite{borg2011assistive}.

Therefore, in our study, we explore the usage and modification of ATs among people with disabilities in China to understand their motivations for modifications, methods, benefits, and challenges.

%developing country残疾人需求非常多，非常重要，而且和developed country很不一样
%一些发展中国家的案例，做了什么调研，发现了什么样的特点
%research gap：发展中国家的残疾人有辅具上的需求，特别是也有很多标准辅具无法满足的需求，但是他们目前是如何解决，面临什么样的挑战，并不清楚，需要在本文中进行调研。

\subsection{Current Public Services for Disabled People in China}
As a developing country, China has the largest population of people with disabilities in the world, totaling 85 million~\footnote{\url{http://factsanddetails.com/china/cat13/sub83/item1906.html}}. Of these, only 37.8 million are registered with CDPF and the National Health Commission (NHC) ~\footnote{\url{https://perma.cc/FX6L-D8UU}}. \hkrevised{According to China's Sixth National Population Census, the three largest categories of disabilities among the Chinese population are physical disabilities (24.72 million people), hearing disabilities (20.54 million people), and visual disabilities (12.63 million people)~\footnote{\url{https://www.cdpf.org.cn/zwgk/zccx/cjrgk/15e9ac67d7124f3fb4a23b7e2ac739aa.htm}}}. After years of economic development and policy improvement, China has established a preliminary public service system for people with disabilities, significantly improving their quality of life and enhancing their social participation~\cite{dong2021study}, however, many scholars believe that this service system for people with disabilities is inadequate in fully safeguarding their rights and promoting the development of disability affairs~\cite{kohrman2005bodies, stone1996law, vaughan1993development,wang2023discussion, lin2019self}. \hkrevised{They raised issues such as governmental bureaucracy, insufficient legal and social support, an imperfect talent cultivation system resulting in a lack of professionals, and the immature development of the assistive technology industry. For professionally trained caregivers, China only has 328,000 individuals, compared to a disabled population of over 80 million~\footnote{\url{https://www.cdpf.org.cn/zwgk/zccx/ndsj/zhsjtj/2022zh/8a6c50f56bae42d9b4cad0440ce01931.htm}}. Also, China has established a multi-level public service system for people with disabilities that includes a "province-city-county" delivery and distribution network. However, the provision of AT in rural areas and local communities remains underdeveloped~\cite{wang2023discussion, zhang2017emergence}. Finally, although China's AT market is rapidly developing and the number of manufacturers is continuously increasing, the supply channels remain limited. There is an oversupply of low-end standardized products, while high-end products are scarce. Moreover, due to varying levels of economic development in different regions, the availability and quality standards of assistive devices also differ, making it difficult for people with disabilities to obtain suitable assistive devices~\cite{wang2023discussion}.} Additionally, in China, people with disabilities are still hard to spot in public places~\cite{campbell2011invisibles} and the majority of them reside in rural areas rather than urban towns, and the educational level of people with disabilities remains relatively low~\cite{li2021choose} and only 27\% of certified people with disabilities are employed~\cite{bruyere2018disability}. 

The AT industry for people with disabilities in China has also made substantial progress over the past 30 years~\footnote{\url{https://www.chinadaily.com.cn/regional/2019-07/25/content_37497824.htm}}. However, there are also challenges. People with disabilities have a low social awareness of ATs~\cite{li2019improving}. A 2015 study by the CDPF found that roughly one-third of the population of people with disabilities required ATs~\cite{jiang2023development}. However, another research indicates that only 7\% of these individuals have access to such devices~\cite{wang2023discussion}. The research also analyzed 138 large rehabilitation assistive enterprises in China and found that the AT industry is fundamentally weak, with an incomplete industrial chain, insufficient variety of AT categories, and low-quality products~\cite{wang2023discussion}.

Additionally, the dire employment situation for persons with disabilities, along with the extra burden of living with a disability, means that economic hardship remains a characteristic feature among the population of people with disabilities in China~\cite{campbell2011invisibles}. Consequently, the AT market cannot rely solely on pricing mechanisms for development~\cite{jiang2023development}.

%数据加到NGOs background里面 NGO background: 没登记也不影响工作流程，要不要讲呢？要讲，但是大幅度精简，主要说明政府主导，管控从严到松，以及慈善捐款数字部分挪下来。外部拿钱很少，资金是大问题。 所有的Context控制在一页。

In addition to commercial enterprises, the provision of ATs for people with disabilities in China is also facilitated through NGOs. NGOs in China can be categorized into two types. The first type is government-led NGOs. The CDPF, established by the Chinese government in 1988, is a representative of this category. Since its inception, the CDPF has been the sole and largest national organization for people with disabilities, with its leadership appointments, funding sources, and decision-making ultimately determined by the government~\cite{zhang2017nothing}. Therefore, the CDPF functions both as an official organization, fulfilling governmental roles, and as a social entity, providing necessary services to people with disabilities~\cite{dong2021study}.

The other category of NGOs consists of those established by persons with disabilities and their families, or by social entrepreneurs. These NGOs are similar to civil society organizations in a Western context. \hkrevised{Due to China's unique political system, social organizations focused on services for people with disabilities find it difficult to obtain funding from the government and are very few in number~\cite{zhang2017nothing}.} Nearly 20 years after the establishment of the CDPF, China's first organization of this type, "One Plus One People with Disabilities' Group", was founded in 2006~\cite{zhang2017nothing, luo2015seeking}. 

In addition, China's immature charity environment has also led to the funding problem of non-government-led NGOs. \hkrevised{Before 1978, the Chinese government held a negative attitude towards the philanthropic sector, leading to the near disappearance of private charity. It was not until the "Regulations on the Management of Foundations" were revised in 2004 that enterprises and individuals were allowed to establish foundations, thereby enabling the development of civil philanthropic activities~\cite{deng2015influence}.}A study by the Harvard Kennedy School Ash Center indicated that in 2014, China's charitable donations as a percentage of GDP were only 0.10\%, compared to 2\% in the United States~\cite{cunningham2015harvard}. However, this landscape is transforming. First, in 2016, a new "Charity Law" eliminated the burden of dual registration for NGOs, established mechanisms for private foundations to raise funds from the public, and promoted transparency, further improving China’s charity environment~\cite{wang2001development}. Second, the demand for NGOs is increasing. Since the CDPF struggles to comprehensively meet the needs of people with disabilities, in many cities, the branches of the CDPF have begun to reform its service system, specifically by outsourcing services to NGOs, which could represent a fundamental transformation in the service structure for people with disabilities in China~\cite{zhao2019changing}.

%Furthermore, according to The Lilly Family School of Philanthropy at Indiana University's Giving USA 2015 report, in 2014, 72\% of total donations in the United States came from individuals, totaling approximately 259 billion\cite{wong2020american}. In contrast, in China, the majority (64.2\%) of charitable donations are made by corporations, and 23.3\% are made by individuals\footnote{\url{https://www.chinadaily.com.cn/a/201809/24/WS5ba82734a310c4cc775e7c10.html}}, which differs significantly from Western developed countries.

\section{Methodology}
% \subsection{Research Context}
% \hkrevised{According to China's Sixth National Population Census, the three largest categories of disabilities among the Chinese population are physical disabilities (24.72 million people), hearing disabilities (20.54 million people), and visual disabilities (12.63 million people)~\footnote{\url{https://www.cdpf.org.cn/zwgk/zccx/cjrgk/15e9ac67d7124f3fb4a23b7e2ac739aa.htm}}. Despite the large disabled population, China has a significant shortage of professionally trained caregivers, with only 328,000 available~\footnote{\url{https://www.cdpf.org.cn/zwgk/zccx/ndsj/zhsjtj/2022zh/8a6c50f56bae42d9b4cad0440ce01931.htm}}. Also, China has established a multi-level public service system for people with disabilities that includes a "province-city-county" delivery and distribution network. However, the provision of assistive technologies (AT) in rural areas and local communities remains underdeveloped. The assistive technology market in China is growing rapidly\cite{jiang2023development}, with an increasing number of manufacturers, but it still suffers from weak in-house research and development capabilities, an immature market, and a shortage of professional talent.

% Also, China has established a multi-level public service system for people with disabilities that includes a "province-city-county" delivery and distribution network. However, the provision of assistive technologies (AT) in rural areas and local communities remains underdeveloped.}

\subsection{Participants}%kexin

This study employs a qualitative research approach using semi-structured interviews. We recruited 20 participants who were involved in the modification process of ATs. We conducted telephone interviews to understand their motivations, roles, processes, and challenges in the AT modification process following the interview guidelines listed in the Appendix.   Interviews took place during time slots when participants had at least one hour of availability. The duration of interviews varied between 50 to 90 minutes, depending on the specific questions and the circumstances of the participants. The interviews were recorded, transcribed, and translated for further analysis. The relevant local ethics review committee approved our study.

\begin{table}[!ht]
\caption{Demographic Information about Demaders}
\begin{tabular*}{\hsize}{@{}@{\extracolsep{\fill}}llllp{3cm}p{4cm}@{}}

\toprule
Participant&Age&Gender&Family Monthly Income &Types of ATs Involved in Modification&Specific Modification Actions\\

\midrule
D1&29&F&Prefer not to say&Prosthetic eye&Prefer not to say\\
\hline
D2&40&M&below 450 US dollars&Electric wheel chair&Modified electric wheelchair 
controller to be positioned near his head\\
\hline
D3&43&M&below 450 US dollars&Electric wheel&Positioned the controller near the chins, installed a phone holder, a stylus holder, a mobile panoramic
camera as a dashcam and wheelchair cushion\\
\hline
D4&59&M&below 450 US dollars&wooden bench&Customized the length, height, and width of the wooden stool, as well as the angle of the legs and adding non-slip features to the bottom\\
\hline
D5&22&M&750 to 1200 US dollars&White cane&Decorated the cane by wrapping it with yarn, covered it with old clothing, and adding stickers for beautification\\
\hline
D6&22&M&750 to 1200 US dollars&Toe straightener&Made rubber toe correctors by using molds\\
\hline
D7&30&M&below 450 US dollars&Toy car for children; electric wheelchair&Replaced the wheels and frame,and added a basket to the car;replaced the battery in the wheelchair\\
\hline
D8&30&F&Prefer not to say&Reading stand&Modified the lid of a cup with a cover to use it as a reading stand\\
\hline
D9&28&M&450 to 750 US dollars&Electric wheelchair&Modified the electric wheelchair to
be foot-controlled\\
\hline
D10&50&M&Prefer not to say&White cane&Added a shelf to the crutch to support the buttocks\\
\bottomrule
\label{DemanderDemographic}
\end{tabular*}

\end{table}
%Demander的表格

\begin{table}[h]
\caption{Demographic Information about Suppliers}
\begin{tabular*}{\hsize}{@{}@{\extracolsep{\fill}}lp{4mm}lp{2cm}p{3cm}p{4cm}@{}}
\toprule
Participant&Age&Gender&Occupation&The Roles in AT Customization&Specific Modification Actions\\

\midrule
S1&34&M&Accessibility Facility Engineer&Manufacturer&Accessibility Environment Retrofit\\
\hline
S2&38&M&Wheelchair Industry&Manufacturer&Electric Wheelchair Modification\\
\hline
S3&21&M&Electrician&Individual technician&Wheelchair Modification Adding Off-Road Wheels\\
\hline
S4&24&M&Software Engineer&Individual technician&Develop English Learning Software for the Visually Impaired, and Enhance the Non-Visual Desktop Access Screen Reader\\
\hline
S5&41&M&Unemployed&Individual technician&Adding a Front Attachment, Modifying the Seat and Installing a Rear Light on the Wheelchair\\
\hline
S6&20&M&Students&Individual technician&Suggestions for Optimizing the Infrared Cane App \\
\hline
S7&40&M&Designer&Nonprofit Organization&AT Design\\
\hline
S8&33&F&Engineer&Individual technician&AT Customization\\
\hline
S9&43&M&Civil Servant&Government&N/A\\
\hline
S10&30&M&Unemployed&Individual technician &Wheelchair and Toilet Chair Modification \\
\bottomrule

\end{tabular*}

\end{table}
%supplier的表格

The 20 participants of this study fall into four distinct categories: 1) Individuals with disabilities who use modified ATs and have been involved in their modification; 2) Individuals with experience in modifying ATs for others; 3) Representatives from third-party modification agencies or NGOs; and 4) Personnel from the accessibility departments of the CDPF. Participants were categorized into two groups based on their roles in the modification process: Suppliers and Demanders. Suppliers are those who provide modification services or ATs for others, while Demanders are individuals who have their own modification needs and receive such services. Detailed demographic information of all participants is presented in Table.~\ref{DemanderDemographic} and Table 2.

\begin{figure}[h]
   \centering
   \includegraphics[width=\linewidth]{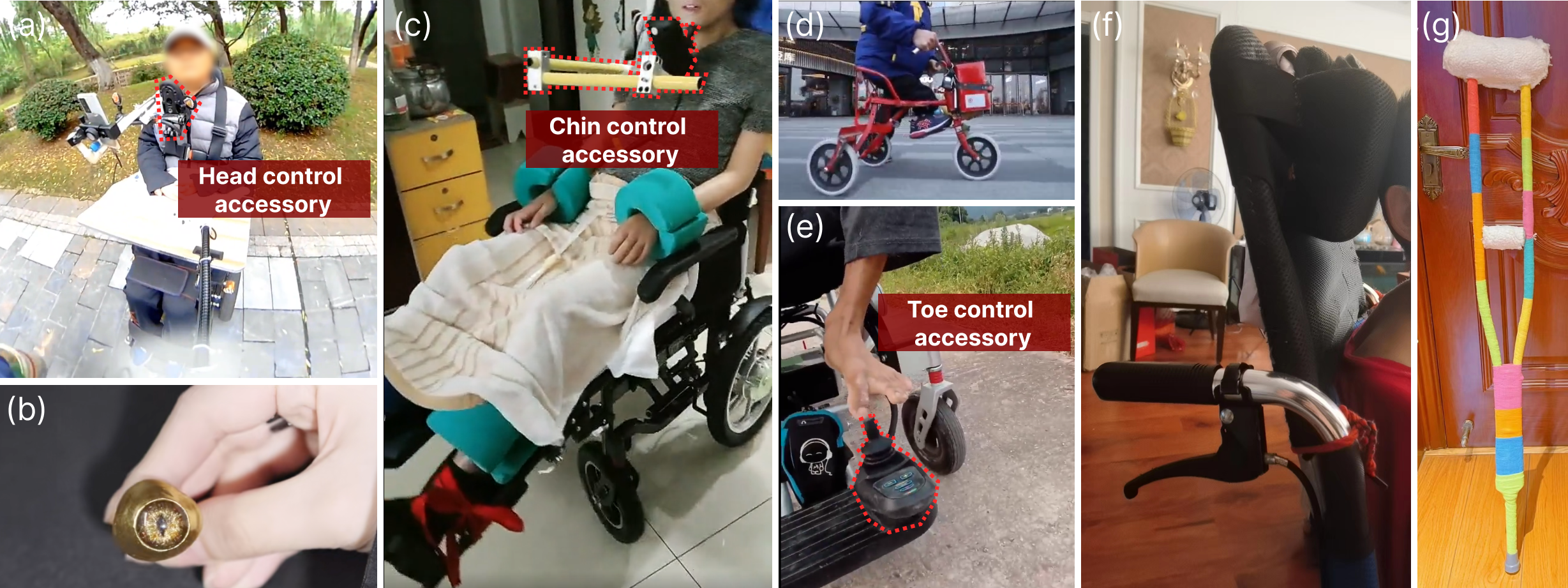}
   \caption{The modified ATs in our study  (The images are provided by participants.): (a) D2’s head-controlled wheelchair; (b) D1’s homemade prosthetic eye; (c) D3's chin controlled electric  wheelchair;
   (d) D7's pocket tricycle;
   (e) Toe-controlled electric wheelchair made by S2 for D9;
   (f) The modified wheelchair backrest S10 installed for his mother; (g) D5‘s handmade crutch appearance. }
   \Description{}
   \label{fig:ATgroup}
\end{figure}

Our initial recruitment involved actively contacting individuals who posted about AT modifications on social media and posting recruitment messages in communities for people with disabilities. This was followed by a snowballing technique, where participants recommended other eligible individuals and shared our recruitment message within their internal communities. Participants received a compensation of 10 US dollars upon completion of the interview.

\subsection{Procedure}%kexin

For the participants representing four different stakeholders, we designed specific questions for the interviews. 

\subsubsection{For Individuals with Disabilities Who Use and Are Involved in the Modification of ATs} We initially inquired about demographic information, their disability status, daily experiences, and experiences with using AT. We then explored the modified ATs they use and inquired about their involvement in the entire modification process. We asked about their motivations for initiating modifications and the physical and psychological impacts of using and initiating these modifications. Additionally, we inquired about their participation in communities for people with disabilities, sources of information and knowledge about ATs, and their expectations for future ATs.

\subsubsection{For Individuals with Experience in Modifying ATs for Others} We first collected demographic information. We learned about their relationship with the people with disabilities for whom they modified devices and the specifics of the modified devices. We explored their entire modification process, including preparation, specific actions taken, and receiving feedback. We aimed to understand their motivations for modifying ATs and the impact this activity had on them. We also inquired about their future intentions regarding participation in AT modification.

\subsubsection{For Representatives from Third-party Modification Agencies and Non-profit Organizations} We began by asking for demographic information. We sought to understand the structure and situation of their organizations, the services and products they offer, and their revenue models. We inquired about their methods of communication with people with disabilities in need and the specific processes of providing services. We also asked about their motivations for starting the service or organization, challenges faced during operations, the current status and reasons for any adjustments or improvements, and examples of failures or successes. We further inquired about their future outlook for services and ATs.

\subsubsection{For Personnel from the Accessibility Departments of the CDPF} We first collected demographic information, followed by a detailed description of their work, the connection between government levels and civil non-profit organizations, and the allocation of financial grants. We inquired about the specific services provided to support individuals with disabilities, methods for understanding the needs of people with disabilities, and the status of AT distribution and modification within government departments. Additionally, we asked about specific challenges encountered in assisting individuals with disabilities and the improvements they hope to achieve in the future.

\subsection{Data Analysis}%kexin

All interviews were conducted in Mandarin by the researcher, whose native language is Chinese. We recorded all interviews with participants and transcribed them. A bottom-up thematic analysis approach~\cite{maguire2017doing} was adopted to identify key themes relevant to our main research question. Three researchers independently performed open-coding on the transcripts. The researchers convened meetings to discuss their codes. In instances of conflicting interpretations, they explained the fundamental principles of their codes to each other and discussed how to resolve these conflicts. Ultimately, a consensus was reached, and a consolidated list of codes was formed. Subsequently, themes were identified from the groups of codes. The results presented in the following section are organized based on these themes, and overlapping themes were clustered together for a coherent presentation of the findings.

\subsection{Research Ethics}
%伦理审查说明
This study was approved by the Institutional Review Board of the authors' institution. Before the experiment, all participants were ensured with informed consent and their right to discontinue the study. Researchers actively assessed the emotional and physical state of participants during the interview process. To preserve participant confidentiality, all personal and confidential information has been anonymized, and the research results presented below have been subjected to de-identification.

%motivation：外部动机都合并到challenge里面， 内部动机继续压缩，不要跟benefit重合。也不能跟related work像，这说明没创新。4.1.4删（跟以往类似，此外）4.1.5 caregiver还比较ok 4.1.6隐藏也删。以往研究已经报告了这些动机。 4.1.7比较独特 2/3页即可。2~3段即可。第一段总结已有的动机，第二段 有趣的发现。benefit语料再典型点，basic need 压缩 语料有点多。 social media导致他们想去DIY第三段-动机来源。 前面强调中国贫富差异大。 淡化义眼的例子，   

\section{Findings}
\subsection{Motivation for Modifying AT}
\begin{figure}[h]
   \centering
   \includegraphics[width=\linewidth]{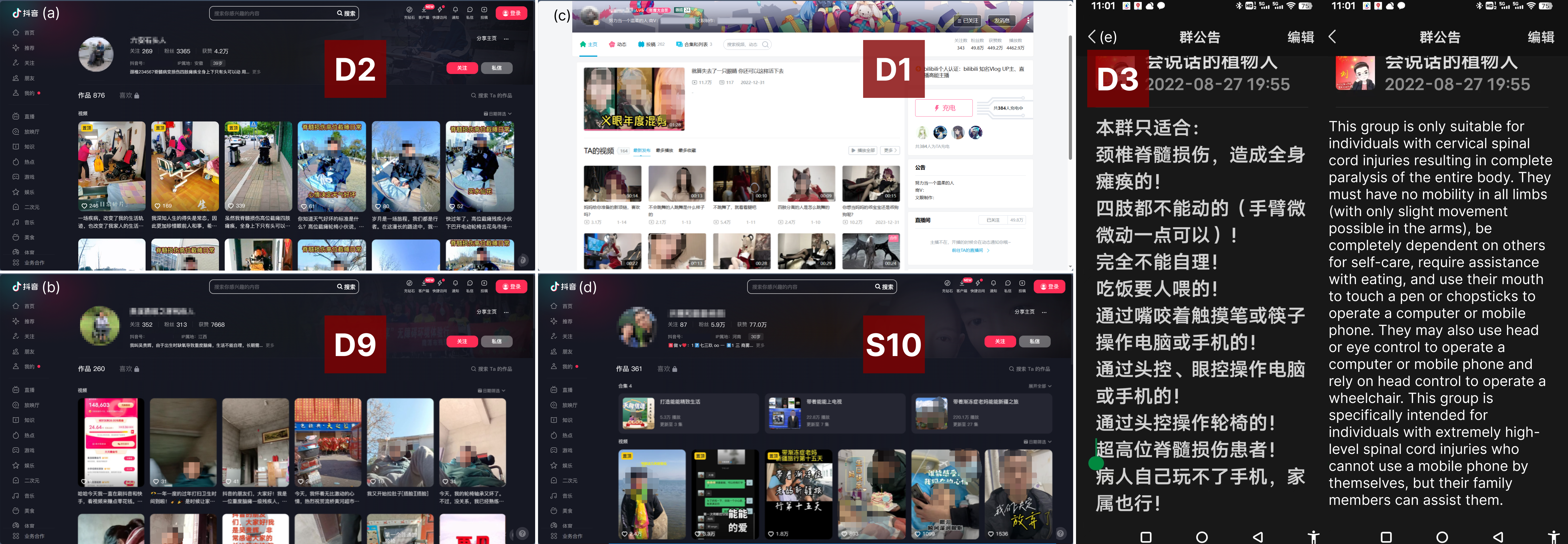}
   \caption{The social media about modified AT in our study: (a) D2's Tiktok account to share modification and rehabilitation knowledge; (b) D9's Tiktok account to share daily life and modification; (c) D1's Bilibili account to share daily life and modification; (d) S10's Tiktok account to share daily life and modification; (e) A WeChat group created by D3 to share daily life and modification.}
   \Description{}
   \label{fig:SNS}
\end{figure}

Some of our participants' reports indicate that their motivations behind the modification of ATs are consistent with previous research.
For example, D2, S10, and D10 reported that standard ATs often fail to meet the personalized needs of individuals with disabilities~\cite{hurst2013making,hurst2011empowering,jacobson2014personalised}; D6 reported that modification allows individuals to conceal their disability~\cite{jacobson2014personalised,shinohara2011shadow,stramondo2019distinction}, and D3, S1 also mentioned that modification helps reduce the caregiving burden on caregivers~\cite{madara2016assistive,nicolson2012impact,mortenson2012assistive}.
In addition to the previously mentioned motivations, Our interviews also revealed internal motivations stemming from personal interest and external motivations stimulated by social media:

\textbf{Personal interest.} Some participants view modifying assistive technology as a personal hobby. S10 explained his attitude towards AT modification, saying: ~\inlinequote{Well, actually, both of us (S10 and his mom) enjoy tinkering with things in our daily lives. It is a hobby we are interested in. So, we often make small modifications to some things at home. For example, we frequently change the cabinets in our house.} What’s more, D1 explained her passion for modifying her prosthetic eye as follows: ~\inlinequote{Because I enjoy creating, I feel it is similar to how some people enjoy photography or painting, except mine is a bit more unique.} Now, D1 is a content creator on the top Chinese video-sharing platform Bilibili~\footnote{\url{https://www.bilibili.com/}} with 400,000 followers, and her beautiful prosthetic eye has become her unique advantage (see Fig.~\ref{fig:SNS} (c)).
%D1-因为喜欢创作，觉得就跟有些人喜欢就是摄影拍照绘画兴趣这样，只不过我的比较特殊。
%S10-其实改装最主要目的就是为了方便，对吧？ 而且其实我俩在平时中d也特别喜欢折腾一些事情，这个是有兴趣爱好的，所以家里的一些东西，我们也会进行一些小的改装，我们家柜子都经常换，怎么说也其实都是来源于生活，就来源于需求。

\textbf{Stimulated by social media.} Additionally, many participants reported that social media inspired their modification behaviors. D2, D5, and D6 mentioned that they were inspired to modify their devices by seeing the DIY possibilities, or others shared their modified ATs on social media. For instance, D2 stated, ~\inlinequote{I came across a video online of someone using a head-operated electric wheelchair, and it inspired me to modify my device along those lines (see Fig.~\ref{fig:ATgroup} (a)).} 
%D2-对，我通过网上面搜索到这个视频以后，我按照他这个视频，就是一个人下巴开着电动轮椅的视频，给我个启发，原来下巴还可以开这栋轮椅，然后我就按照这个思路去自己想来，按照我这个想法。
%Furthermore, social media, TikTok~\footnote{\url{https://www.tiktok.com/en/}} specifically, has also stimulated motivation for AT modification among the providers of these modifications. For instance, 
S8 learned about D9 through social media (TikTok~\footnote{\url{https://www.tiktok.com/en/}} specifically) and, after communicating with them, customized an electric wheelchair for D9 based on their specific needs, providing it for free.

Overall, we found that the main motivations for Chinese people with disabilities to customize and modify AT currently include meeting personalized needs, concealing their disability, reducing the burden on caregivers, treating modification as a hobby, and inspiration from social media for modification ideas.

\subsection{AT Modification Process}

\begin{figure}[h]
   \centering
   \includegraphics[width=\linewidth]{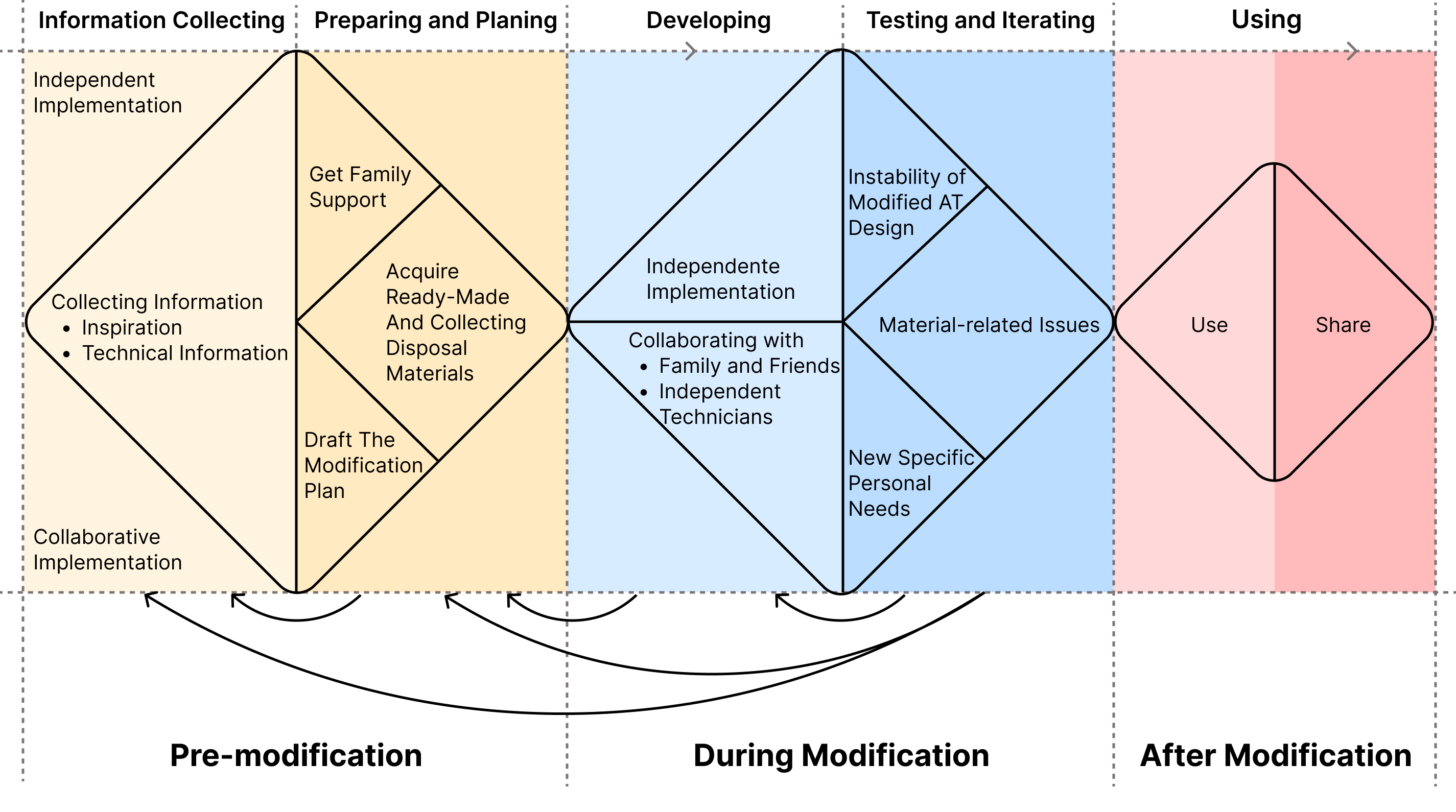}
   \caption{The modification process (adapted from the Double Diamond design process model~\cite{council2005double}).}
   \Description{}
   \label{fig:modification_process}
\end{figure}

 %\includegraphics[width=1\textwidth]{fig/Modification process.png}
% Our \joeyrevised{interview findings} results \joeyrevised{indicate} that
 From our participants' report, aside from a minority of people with disabilities who can find relatively professional organizations to help modify AT (this part will be detailed in Section~\ref{BMSec}), most participants still need to undertake AT modifications mainly by themselves or collaboratively, such as family and friends who are capable of modification skills, and technicians willing to provide help. 
 Thus, the modification process often involves \joeyrevised{social networks around individuals with disabilities.} Modifiers include those led by people with disabilities or cohabitants. 
\joeyrevised{Due to the lack of a unified workflow in the modification process, we found that based on the results of open coding, users often act as (co-)designers in completing the modification of assistive devices. To better describe this process, we attempted to use the double diamond model in design to outline their workflow. (see Fig.~\ref{fig:modification_process} ).}
 It mainly includes the following stages: collecting information, preparing, implementing, testing and prototyping, using, and sharing the modified program.

\subsubsection{Information Collecting Stage}
In this stage, individuals with disabilities seek modification inspiration from social media and market references, tap into personal and public sources for technical knowledge, and identify capable organizations for AT customization before modification.

Individuals with disabilities or their collaborative modifiers gather inspiration on how to build ideas in various ways before formal modification. They mainly learn their skills from craftspeople who are online bloggers. For example, S10, D6, and D2 watched some of the videos posted by crafty people or other people with disabilities on the SNS (Social Network Services) platform. They find inspiration in these handmade techniques and modification processes. D6 said: ~\inlinequote{Because some people on the SNS platform have the same symptoms as me, and it is useful for me to see how they go about doing that stuff.} 
%D6-之前的话可能说在网上他不是想抖音上有就像那些手工人那些的话，我感觉他们做的一些可能说能够对我有一些帮助，因为像市面上买的那些都太贵，然后找着设计师而言，价格太离谱了，有时候就会去看别人，然后做的那些东西。
%D2-我通过网上面搜索到这个视频以后，我按照他这个视频，就是一个人下巴开着电动轮椅的视频，给我个启发，原来下巴还可以开这栋轮椅，然后我就按照这个思路去自己想来，按照我这个想法。” 
%Furthermore, Modifiers who make appearance modifications will combine with personal aesthetics. D1 said,\inlinequote{Some of the inspirations are from games and anime and stuff like that, and then some of them are landscapes or mountain ink drawings (see Fig.~\ref{fig:ATgroup} (b)).}  %Some people with disabilities obtain technical information based on their interpersonal resources and some public knowledge. D1 researched foreign high-tech prosthetic eye information through the Internet and the hospital’s eye-related information literature.

Furthermore, the participants utilize a diverse array of sources, ranging from social media to professional networks, to gather the necessary inspiration and technical knowledge for effective ATs customization. D1 researched foreign high-tech prosthetic eye information through the Internet and the hospital’s eye-related information literature.
%D1-很多灵感有些是游戏动漫之类的，然后有些是风景或者山水墨画这些都可以成为灵感。
%D1-我有两个合作的义眼师，但他们只会做普通的义眼，然后是可能会参考一下他们的意见；也会通过网络查阅国外高技术义眼信息，北京同仁医院关于眼睛的相关文献。也会联系国外的义眼师，但是咨询的次数没有特别多。 但具体来说还是我靠自己不断的试错。”

%Looking for organizations with modification capabilities
% Individuals with disabilities opting for AT modifications initially turn to organizations that offer such services. As D2 said, she tried several times, including searching on the internet and online community forums, seeking organizations in this area.
% %D2-对，我尝试过好几次，包括在网上面搜索，包括我以前在贴吧里面发帖子，寻求这方面的机构，都是没有。
% Due to limited knowledge, some people with disabilities are not aware of the existence of modification organizations. D3 and S10 have similar experiences. For instance, D3 mentioned: \inlinequote{Because of the limited knowledge in the beginning, I do not know that there are modification organizations} 
%D3-因为刚开始认知有限，不知道有改装机构或者说是已经改好的时候，不去包括现在我都是还是想自己改装，因为也根据自己能力有限，你像国外听说有改好了的那种功能比较多的，但是也买不起，是对于我这样的，大多数我这样的患者是买不起，就是少部分条件好的肯定是买得起的。

\subsubsection{Preparing and Planning Stage}
Generally speaking, after collecting enough information, people with disabilities or their collaborative modifiers who conduct modifications will start to prepare materials and plan the draft modification designs. Importantly, in this stage, people with disabilities who live with others, such as their families, need to obtain their cohabitants' permission.

Most people with disabilities who cooperate in completing modifications would obtain family members' permission. Especially for D3, the importance of family support is deeply felt. He said: ~\inlinequote{I showed my family a man driving a wheelchair with his head and said I could do the same since my head functions normally. Being in this field, I am confident in continuously communicating with my family when they are receptive.} S1 also pointed out that family member’s support was important, He said that some of the family was dissatisfied with the renovated facilities, as the addition of a handrail clashed with their decor style and was not aesthetically pleasing, leading to a redo.
%D3-刚开始给家人看了一下，这个人可以用头开轮椅的，我说我一定也能可以，因为我知道我的头是正常的，跟他一样。而且我是干这行的，是有把握，家里人心情好的时候就跟家里人不断交流。
%d2-首先我改装这方面是得到家人的支持，家人的支持这方面只有家人支持我才可以出门，没有家人支持我出不了门，没有家人支持我也改装不了电动轮椅。也就是说我今天能心态变好，这也是家人的支持，但是外界的人我不知道什么看法，但是我在外面的时候我也遇到我感觉是不太好的声音。

After obtaining support from their family, some people with disabilities purchase ready-made materials or collect discarded materials within their living circles. D6, D7, and D8 purchased materials from online platforms. 
%like D7 said: ~\inlinequote{I browsed and chose items on Taobao~\footnote{\url{https://world.taobao.com}}, and then consulted Taobao customer service about the material polyurethane.} 
D2, D5, and D7 have similar experiences in collecting disposal materials. D7 had to collect materials from scrap collection sites since the original wheelchair was no longer for sale. He said: ~\inlinequote{I got the skeleton of this car from a scrapyard. Although I can only find this wheelchair on Taobao, its carrying capacity and quality are very poor (see Fig.~\ref{fig:ATgroup} (d)).} 
%D7-后来在某宝上面我就挨家挨户的在看，在选，然后就问人咨询人家在淘宝上对淘宝这样聚氨酯的。
%D7-这个车的骨架是我从废品收购站我淘回来的，现在已经买市场上买不到这个车了，现在在淘宝上面能收到80 90的回忆，这个名称能收到这个车，但是它的样式大概一样，但是它的承载力各方面，质量是很差劲的，就是一坐会坐散掉的感觉。
%D2-第一个坐垫的海绵是家里面自己做的。从从小区里有很多装修的拆下来的海面，就是自己捡回来，然后缝合一个布套。
A few people with disabilities will draw a draft before implementing it, such as D7 and D2. D2 completed the drafting of the modification sketches on the phone using a capacitive pen.

This stage is an effective transition from idea to execution for individuals with disabilities and their collaborators. It involves not only collecting materials and drafting plans but also obtaining the support and approval of family members.
%D7-我把尺寸精度把握好，然后图纸画好，基本上没有什么困难。
% 这儿还可以有电容笔的语料去支持Draft，但是会和challenge地方的语料重叠，所以draft这个内容还需要吗？因为图里面有Draft的这个步骤，或者说把这个draf在图里面删掉？
 
% \paragraph{\textbf{\textit{(3) Get Family surpport}}}
% Most people with disabilities who cooperate in completing modifications would obtain family members' permission. Especially for D3, the importance of family support is deeply felt to him. He said: ~\inlinequote{I showed my family a man driving a wheelchair with his head and said I could do the same since my head functions normally. Being in this field, I'm confident in continuously communicating with my family when they're receptive}
% %D3-刚开始给家人看了一下，这个人可以用头开轮椅的，我说我一定也能可以，因为我知道我的头是正常的，跟他一样。而且我是干这行的，是有把握，家里人心情好的时候就跟家里人不断交流。

% S1 pointed out that family member's support was important, He said that some of the family was dissatisfied with the renovated facilities, as the addition of a handrail clashed with their decor style and was not aesthetically pleasing, leading to a redo. 
 
 \subsubsection{Developing Stage}
 During the installation process of ATs, some people with disabilities can complete it independently, while most participants need to collaborate with others to complete the modifications.

 %Some people with disabilities independently complete their modifications. 
 For some simple modifications, capable individuals with disabilities can independently complete the alterations.
 For example, D8 and D5 did a simple modification to ATs by themselves. D8 attached magnets and magnetic plates to a water bottle lid, allowing a phone to be fixed onto the lid, thus creating a phone stand to help her read.
 %D8-引磁片它本身是没有磁性的，就有个背胶，然后可以贴可以贴上去，首先是引磁片稳定住了，然后磁铁才可以在中间发生作用，内就引磁片贴上去的时候，才可以算是完成改装
 D5 wrapped an item with foam padding, similar to that used in packaging. Additionally, he used yarn for colorful, wrapping in some areas and repurposed an old piece of clothing for extra comfort at the top, near the armpit area (see Fig.~\ref{fig:ATgroup} (g)). 
 %D5-这个里面我把干的那个地方就是包裹了一层泡沫，泡棉就是那种快递里面那种泡棉，然后不是那种给它包裹了一圈，包裹了一圈了之后，然后就用那一些又然后下面的这种一些颜色的话，就是用毛线弄的，就是一一一一圈的鼓鼓裹，然后最上面的枕着胳肢窝这个地方的话，用的一件衣服就是不用的衣服，然后进行一个给它裹住.
%删掉4.26-D1 emphasized full participation in the process by saying, ~\inlinequote{no one can help me with things like craftsmanship or technical stuff. Constant trial and error, just keep trying on my own, The information on the Internet can only be used as a reference because the production processes are different.}
 %D1-像工艺或者技术类的这些东西是没有人可以帮到我的。不断的试错，就自己不停的试。网上的资料只能作为参考，因为是不同的制作工艺。
 
%\paragraph{\textbf{\textit{(2) Collaborating with Family and Friends}}}
In the complex collaborative modification process, their family members, friends, and independent technicians mainly handle operations, procure materials, and collaborate with people with disabilities, ensuring physical installation and communication needs are met.

As some people with disabilities are immobile, family members need to be responsible for the \joeyrevised{communication,} installation and material collection. \joeyrevised{D2 and D10 have modified the controller rod and backrests of their wheelchairs. The installation of the controller rod for D2 was done by family members. D2 also described: ~\inlinequote{My dad uses various clamps that I bought online to assemble steel and water pipes.}}
%D2-轮椅我基本上是改过几次，我来看一下。第一次的时候我就没有想到找电焊师傅，第一次的时候我就是从网上面买了各种各样的管夹，夹具，自己回来，然后用钢管自来水管，用夹具各方面拼起来的，当时用也可以，只能在家里面，比如说路面一颠簸的可能就不行了，毕竟那些夹具是夹不住的，最后我才想到找电焊师傅，按照我的方法去焊，焊一个就是固定控制器的支架，包括家具，这样改装以后效果就挺好。
Including the user's ideas in the design process is a prerequisite for effective modification. 
S10 mentioned:
    ~\inlinequote{My mother and I communicate frequently, and she also has a lot of ideas. For example, when we need something, which thing can be replaced, and how to get it. After we communicate like this, we will know.}
 \joeyrevised{To address the issue of uncomfortable and unstable backrests, S10’s mother played a significant role in providing methods to S10. She instructed S10 to remove the cushion from the lower leg area of the wheelchair and sew it onto the backrest which bought on the internet, creating a small hole in the excess part at the back. Then, the high backrest was inserted into it. (see Fig.~\ref{fig:ATgroup} (f)).}
%S10-我妈妈我俩会经常沟通，然后他的点子也特别多，他虽然所以这个病会让他越来越聪明，他可能就坐下来静看世界的情况下，我们家任何一个东西她都想到，我们俩比如说需要一个这个东西，哪个东西可以替代，怎么样搞上去，我们俩这么一沟通一交流，这么就成了。”

%\paragraph{\textbf{\textit{(3) Collaborating with Independent Technicians}}}
D2, D3, D7, and S10 would look for independent technicians to help with modifications. \joeyrevised{For example, D7 initially consulted with an electric scooter shop owner and decided to replace the lithium battery in his electric wheelchair with a lead-acid one, which required a new battery box. When the box could not be installed in its original position, D7 worked with a welder to devise a solution. They considered welding an iron bracket to mount the box, but D7 preferred a flexible approach to maintain the wheelchair's foldability. 
%\inlinequote{The welder was originally going to weld it directly, but I said no, the bracket on the wheelchair must be movable because it would not fold otherwise, right?} He explained. 
Eventually, the welder proposed using screws as a fixation method. By drilling holes in the bracket and then inserting screws and pins, the battery box was securely mounted on the back of the wheelchair.}
%D7-然后现在的话，如果这个锂电池我也不怎么用，我要花1000块钱再换一个锂电池，我拿不拿不出这么多的钱知道吧？他是24伏的，后来我就想了一下，跟卖电动车的老板想了一下，我说在网上买了一个电动轮椅的装铅酸电池的电池壳，两块电池，因为它电池也没有电池壳固定，你也一个不安全，第二个充电也不方便对吧？然后改装了一个电池盒，电池盒弄好了以后，然后问题来了，说啥问题？它电池没办法固定，你放在左边也不行，放在右边又不行，不是锂电池的轮椅，它的电池是放在轮椅的后面，屁股后面知道吧？但是我他也不是那种的轮椅，他没办法去固定。后来我就就想了一下，我用铁，把轮椅的电池要挂在我轮椅的后面，要不然的话他也走不起来，知道吧？后来我就自己设计，但是没画图纸，你要图纸我哪天我要画一个拍给你看一下。凭我自己的经验，我就教人家电焊师傅，就是我让他怎么弄，把尺寸精度把握好，师傅后来本来是准备把它电焊上去的，我说不行，挂轮椅的铁架一定要用那个活动的，因为它折叠起来的时候，你看上去轮椅它就没办法折叠了，对吧？然后就通过在轮椅上面准备电焊，然后用螺丝作为固定把它勾上去知道吧？但是它这个是属于那种钛合金的那种材质的，电焊又不好焊，头疼了，没办法，后来师傅说你就放在我这边，我晚上给你弄，然后用电钻打眼进去以后把螺丝再焊在螺丝上面，然后通过两边的销子把它固定在挂在上面，然后现在轮椅的正常使用，它就是锂电池拿1000又重了大概十几斤左右增加了负重，但它的行驶里程也缩短了，因为它没有锂电池那么轻对吧？改装了一下，花了大概200多块钱，200多块钱。换了一个电动车上面的就是12古代的一个电池换了两块，两块的话在淘宝网上买了一个电池盒，电池盒是45块钱，然后电池是90块钱。因为我拿两个旧的给他换的知道吧？然后那个挂架，焊的挂架，本来是准备网上有现成的挂架，但是尺寸也不够，然后只能用铁把它挖一些凹槽出来。挂架就叫人家连材料带电焊，花了50块钱，就这么多钱，后来我就用银粉漆把它在家里把它喷了一下，跟轮椅的漆的颜色是一样的，他要不然会比较难会生锈的，知道吧？
D3 and D7 found the right technicians through previous work relationships. For instance, \joeyrevised{D3 reached out to a former colleague, scheduled a time, and instructed him on how to make the modifications. %He said, \inlinequote{Since we are in the same field, it was easy to explain, and then I sat in it to do the adjustments.}
} 

People with disabilities either undertake modifications independently or collaborate with their social networks. They often use everyday tools and materials for simple modifications, as well as engage in complex modification processes that involve technical tools and multi-party communication.
%D3-我找以前认识的上班的工作的同事，然后我零件买好了，预约时间叫他看有时间方便来家里帮我我把我的需求告诉他，因为是同行人，一说就懂了。他就按照我的要求改到八九不离十之后，然后我人坐上去之后他再帮我调试，然后调试” 

%TODO:应有总结句 残疾人或独立实施改装，或与他们的社交网络合作。其中包括针对日常活动的简单改装和涉及专业技术的复杂改装。

\subsubsection{Testing and Iterating Stage}~\label{testSec}
Following the completion of the developing stage, individuals with disabilities will start to test the modified AT and often will encounter several issues, including instability of the modified AT Design, material-related issues, and new specific personal needs, leading to subsequent iterations.

%\paragraph{\textbf{\textit{(1) Instability of Modified AT Design.}}}
Modified ATs can face instability during use. When D2 used his head to control the modified wheelchair outdoors, he found that the controller would become loose due to the bumpy road. To solve this problem, D2 added support rods (see Fig.~\ref{fig:ATgroup} (a)). 
\joeyrevised{D2 visited the store and shared his ideas with the hardware store owner three times, hoping the owner could help him. Ultimately, the shop owner helped resolve the stability issue.%Finally, The owner then measured the dimensions and distances, began the specific operations, and installed the rod with minimal talking. To further strengthen the stability of the high backrest of the wheelchair in the iteration, S10 removes the string from a gift box and ties it around the backrest, adjusting the angle of the high backrest by the tightness of the string. He said,\inlinequote{The idea of using the rope was my mom's because it was always unstable. She saw two carrying ropes on the gift box and suggested that I take them off and use them to secure it.}
}
People with disabilities discovered material defects after using modified ATs, so they need to iterate the modification design. For example, D2 discovered the issue of poor fit with his cushion as he tends to slide down easily while sitting in a wheelchair with his deteriorating back and waist muscles, he began to gather information anew and discovered that he could buy the desired cushion online (see Fig.~\ref{fig:ATgroup} (a)).
However, due to the poor durability of the material. Thus, D2 collected information once again and found the car seat cushions from the dismantled car factory to match his needs. \joeyrevised{His father went to the car scrapyard on his behalf and helped him measure and buy the cushion.}
%He described: ~\inlinequote{Finally, now I have replaced it with a car seat cushion. The real car seat cushion is the car rental one I found from a car dismantling factory. I put it on and it feels very good now.}
%D2-最后现在就是换成了汽车坐垫，真正的汽车坐垫，就是从拆汽车的拆旧厂里面找的汽车租赁，装在上面现在感觉用的挺好  

%省字数，不要了-D7 also mentioned that he could not repair the wheel due to the special nature of the wheel material after he used the modified tricycle for a while. He got inspiration from other assistive parts and came up with new modification ideas and bought a more expensive PVC wheel on the internet and replaced it.
%D7-后来轮子坏了,也买不到一样的了，修也不好修，然后他铁又比较薄，电焊也不能焊，没法焊住。焊了之后过几天，它折腾一下还是断了。我就突然想起来，我说干脆用之前买的轮椅的前轮，头一次买来用了大概三年， Pvc质量还挺好的。然后去年年底的时候，在淘宝上买了一个较贵的PVC轮子。

%\paragraph{\textbf{\textit{(3) New Specific Personal Needs.}}}
\joeyrevised{Some people with disabilities often develop further needs after using ATs, preferring additions that enhance functionality, entertainment, safety, noise reduction, and labor-saving.} In case of an emergency, D3 wanted to make emergency calls from his mobile phone when he was out on his own, so he fitted a mobile phone holder to his modified wheelchair (see Fig.~\ref{fig:ATgroup} (c)). \joeyrevised{Since S10 often takes disabled individuals out and their home lacks an elevator, lifting his mother's wheelchair down the stairs due to her muscle atrophy is strenuous and requires at least two people. To resolve this, S10 undertook environmental modifications by installing a ramp outside their residence.}

\joeyrevised{Throughout this phase, individuals with disabilities actively engage in testing and reworking their modifications, addressing issues related to design stability, material durability, and evolving personal needs.} 

\subsubsection{Using Stage}
Some people with disabilities are satisfied with the functions after using the modified ATs, Like S10, D2, D3, D4, D5, and D7. 
%S10 believed that modified wheelchairs that meet the needs of users would not pursue high-priced and high-performance wheelchairs. 
%不要"I wanted to buy the Anwei wheelchair and have a try, but my mother was unwilling to use it because the wheelchair was already very good and she did not want to spend more than 10,000 yuan on it. "
%S10-安维轮椅当时确实想买，也想试一试，然后我妈就不愿意用，因为现在轮椅已经做了已经很不错，再花1万多买，他也有点不舍得。
D7 was very satisfied since the modification was completed. He said: ~\inlinequote{I have no dissatisfaction, because my little tricycle has been modified for 10 years, and I do not think there is any inconvenience (Fig.~\ref{fig:ATgroup} (d)).}
%D7-我没有不满意的地方，因为我这个车改了10年了，我觉得没有什么不方便的。

%\subsubsection{\textbf{Share}}
When people with disabilities are satisfied with using modified ATs, many of them(D1, D2, D3, D7, D9, S10) would share their modification experience on social platforms, S10 (Fig.~\ref{fig:SNS} (d)) and D3 (Fig.~\ref{fig:SNS} (b e)) share their modification methods and sources of purchasing accessories on social platforms and communities. D1 (Fig.~\ref{fig:SNS} (c)), D2 (Fig.~\ref{fig:SNS} (a)), D7, and D9 shared the ATs they modified for everyday use on social media (Fig.~\ref{fig:SNS} (b)). 
%S10 said, "In fact, many people came to ask me about the content I posted on Tiktok. If I could make these accessories, they would definitely be willing to pay. Then my general way of dealing with it is, first, if I bought it, I will recommend the relevant links to them, and if I did it myself, I will tell them the whole process." 
%S10-其实很多人过来，你要说我有能做出来这些配件，他们肯定是都愿意付钱的。然后我的一般的处理方式是，第一，如果说是我买的，我会把相关链接推荐给他们，如果是我自己做的，我就会把整个流程整个需要用它互做给他说过去”。
%语料和改装好处关系链接有重叠-不要-The modification method is effective when similar patients use D2 to share it, allowing more people to complete the modification and also increasing the possibility of manufacturers accepting customization. D2 mentioned, "Then they will modify it according to my method, It feels pretty good to use, and they will also find manufacturers to customize it according to my method."
%D2-然后他们会按照我这个方法去改装，也有按照我这个方法找厂家定制，就直接找厂家定制，也有这样子的。“怎么说？只能说轮椅改装了以后给我的生活论语生活带来了很多的方便。我也把我这个方法告诉我，和我一样级别的情况差不多的，伤友们也有好几个伤友，根据我这个要求，然后也找人改装，使用感觉还都挺好。

Overall, the process of modified AT includes collecting information, preparing and planning, developing, testing and prototyping, and using stages. 
%In these stages of modification, there are instances of backtracking before moving on to the next step. Modifiers include those led by people with disabilities or cohabitants. 
Throughout this entire process, multiple stakeholders are involved, such as people with disabilities, cohabitants, family, friends, and independent technicians. Modifiers often collaborate closely with multi-stakeholders, each contributing their expertise and ideas to enhance the effectiveness of the AT.

\subsection{Benefits from Modified AT}
This section focuses on the multidimensional benefits of modifying ATs for people with disabilities. Through in-depth studies, we reveal how modifications enhance not only practicality but also greatly improve the psychological health and quality of life of users. Therefore, the benefits of modification for ATs not only include improvements in basic needs such as comfort but also enhancements in psychological, social, and self-fulfillment aspects.
%These enhancements not only improve the practical functionality of the devices but also significantly impact the emotional well-being and personal development of their users, particularly those with disabilities. Each category encompasses unique advantages that together contribute to an improvement in their quality of life.

\subsubsection{Enhancing the Comfort and Independent Living Ability}
People with disabilities often endure a lot of discomfort while using their original ATs. Modifying these devices can greatly enhance their comfort and reduce their pain. %D6 replaced AT material to relieve the pain of the big toe. He mentioned: ~\inlinequote{Previously, the bone next to my big toe often hurt. It became less painful after I replaced plastic with rubber material myself.}
%D6-他提到，“以前是用的塑料，大拇指旁边那块骨头就经常疼，那时候经常用的话就非常疼。后面就可能说使用的话就没有那么疼。”
%Due to the special conditions of D7, he cannot sit on the seat of a small tricycle. He can only sit on the backrest of the seat, which is a thin pillar， He cannot sit in his seat for a long time as his buttocks get numb. He tied tape to the thin pillar to prevent damage to the buttocks.
%屁股角度会发麻，因为它就是一一个柱子，圆柱子在上面对不对？它没有座椅，你实际上刚才在上面它肯定会发麻的，对不对？
S10 had a cushion modified and improved her mother's comfort, as she would have endured a lot of discomfort in daily life.

%为了精简，删掉了第三个语料：S10 usually thinks about how to improve the mother's comfort by modifying ATs as she would have endured a lot of discomfort in daily life. Considering the discomfort caused by the unique symptoms of people with disabilities when using ATs, S10 had a wheelchair modified. He shared, ~\inlinequote{Due to the muscle atrophy in her buttocks, leaving only two large bones, I considered digging a hole in the cushion. This way, by reducing the pressure on the area, the pain was significantly lessened, and she felt more comfortable."} 
%S10-最简单就是它坐垫，换了好几种很多材质坐垫，然后不行，后来想着用乳胶枕，他因为他屁股骨头肌肉萎缩，就剩两个大骨头了，就想着能不能把要挖个洞，它接着做进去这样它受力减少的话它就没那么疼了

%------独立
%除了获得辅具的舒适感，无法行动的残疾人向往去住处的外面。改装扩大了无法行动的残疾人的活动范围，他们能够独自去到其他场景中。
In addition to improving comfort, AT modifications can also increase the independence of people with disabilities by allowing them to independently visit different locations. D2, D3, and D7 all mentioned expanding the scope of activities. After modifying the wheelchair, D7 can not only move freely in the local area but also take the wheelchair to other cities ~\joeyrevised{with highspeed rail}. D7 said: ~\inlinequote{If I go out of town or to any other city, I carry this with me.} ~\joeyrevised{D7 also found out how to take his small modified tricycle on a plane to other places. He said, ~\inlinequote{When flying, my vehicle needs to be packaged and checked in. I have heard that people with disabilities need to notify the airline in advance if they are bringing a wheelchair; my friends have done it several times.}}
%D8-如果说我去外地的话，或者去什么其他城市的话，这个车都带着的随身带放在大巴肚子里面。对于P3
%这句放在4.3.1不合适4-26：He valued his modified tricycle as much as important as his life. He told us:~\begin{quote}  ~\inlinequote{It is my legs, my life; I take it wherever I go (see Fig.~\ref{fig:ATgroup} (d)).}\end{quote}
%D7-他就是我的腿，就是我的生命，就是我到哪去，我都要带它。除非去哪里玩了，然后不需要用脚，就会带个带手推的轮椅，然后可以有志愿者推或者说家人推。

%For D2 and D3, before the modification, due to physical limitations, they completely relied on family members to push their wheelchairs when going out. D2 mentioned, ~\inlinequote{This change was very significant. If I did not modify my wheelchair, I could only go out if my family was available to push me. The modification allowed me to live more independently.}
%D2-这个改变是非常大的，改装之前，你像我这样自己身体不能动，自己身体不能动，自己生活不能自理，一个可以说每一个人都希望自己不麻不麻烦别人。如果我不改装轮椅的话，，那么我想出门必须家人受推出门，必须在家人方便的情况下我才可以住。

%Although some of people with disabilities still need assistance from family members to get into the wheelchair, once they are in the wheelchair, they can go out independently. Like D2 and D3, D2 could go out on his own once his family helped him into the wheelchair after modification. D3 also said, ~\begin{quote}
    %~\inlinequote{Now, I can stroll around by myself within a 5-kilometer radius.} 
    %\end{quote}
%D3-后来就尝试着一个人出门附近转转，他说， “现在5公里范围内都可以一个人去溜达，没问题。”

Modified AT significantly enhances comfort and independence for people with disabilities, alleviating discomfort and enabling greater mobility.

%与家人关系改善
\subsubsection{Reducing the Burden on Caregivers and  Improving Family Relationships}

The modification of ATs has brought considerable convenience to caregivers and patients, particularly in reducing the workload of caregivers, thus improving the relationship between them. ~\joeyrevised{For S10, after his mother underwent surgery in the early stages, she could only urinate and defecate in bed. S10 said, ~\inlinequote{Caring for a bedridden patient puts me on constant alert and brings significant difficulties.} S10 welded a frame similar to a toilet seat, allowing his mother to use the restroom alone, which also reduced his caregiving workload and preserved dignity for both parties. 
%He explained,~\inlinequote{Being cared for in bed for bathroom needs is undignified, and it is the same for the family members.} Furthermore, S10 and his family were confined to their home before modification, unable to go anywhere. After the modification, S10 and his family can now take their mother out together. As he said, ~\inlinequote{We push the wheelchair up the mountain so all of us can enjoy the natural scenery and breath the fresh air, instead of us all staying at home every day.}
}

D2 also mentioned that after modifying the wheelchair, he no longer needed family accompaniment, which greatly reduced the hassle for his family (see Fig.~\ref{fig:ATgroup} (a)).
Modified ATs ~\joeyrevised{have simplified the routine of daily care and created opportunities for outings together, providing caregivers and recipients with an environment to enhance family relationships.}

\subsubsection{Shifting Attention Away from the Disability Itself and Alleviating Social Pressure.}
Some modified personalized ATs can help people with disabilities relieve social pressure, especially those with unique appearances. When people with disabilities are out and about, the attention of those around them is diverted from the person with a disability to their unique, personalized ATs. Therefore, these unique ATs serve not only functional purposes but also act as tools to shield against societal pressures. ~\joeyrevised{Previously, D5 felt timid when going out, afraid of other people's stares.} He told us: ~\begin{quote}
    \inlinequote{I went out with my modified colorful crutches. People first noticed my crutches, not my body, shifting their attention to the device. It is like the device shields me from negative remarks (see Fig.~\ref{fig:ATgroup} (g)).~\joeyrevised{Some passersby would compliment me, and I would feel a bit proud.}}
\end{quote} 
%D5-然后还有用了这个东西之后，也会觉得让自己更加阳光了一些，原来还是比较消沉，用了这个之后走到外面的话也不会一些很胆怯，或者是走到外面会有一种特别被别人指指点点这样的一些说法。但是用了之后的话会有一种没有这种一些想法了，就会觉得他们的一一第一眼就看的是拐杖，不是看我这个人的一些体就是那种身体，会觉得相当于会把他们的一个目光然后转移到了他身上，而也相当于让他给我抵御了一些那种不好的一些言论什么的。

%The modification of ATs significantly reduced the social pressure on D6. Before the modifications, D6 was fearful of others discovering his disability and made efforts to hide it. He wore shoes that were larger than his size and frequently trimmed his toenails. D6 expressed concern about being noticed while browsing ATs on the internet. He stated: ~\inlinequote{When I browse online for ATs related to my condition, I worry about being noticed by others buying such things. I am afraid of being seen by many people around me, which creates a certain anxiety in me.}
%D6-在网上浏览和自己相关辅具的时候，会担心被别人注意到自己在购买这样子的东西，我怕被身边好多人看到，会存在这种心理，”
Additionally, the modified ATs widen the activity range for people with disabilities and offer them opportunities to engage with the outside world, thereby helping them to gradually overcome social anxiety.
%另外，能够拓宽残疾人活动范围的辅具改装，给予了他们能与外界接触的机会，逐渐克服社交恐惧。
D2 shared that he overcame his social fears and alleviated his sense of social disconnection by using a modified wheelchair, gradually increasing the frequency of going out, and eventually eliminating his worries about going out. ~\joeyrevised{Before the modifications, D2 rarely went out and felt completely disconnected from society. He said, ~\inlinequote{I feel extremely socially anxious, always avoiding people when I see them.} Since the wheelchair was modified, it has attracted the attention of onlookers who thought he was driving a high-tech wheelchair and approached to ask about it. %Initially, he would avoid them and feel nervous during these interactions. 
Now, P3 can proactively explain his modifications and the reasons behind them to the spectators, ~\inlinequote{Now, people in the park generally recognize me and know that I modified a regular wheelchair myself}, he explained.} 
%D2-刚出门的时候，整个人感觉与社会已经完全脱节了，感觉整个是人是非常社恐的，看见人都是躲着的，但是经过轮椅改，但是轮椅改完过以后，出去一次两次三次、1月两个月一年，现在我可以这么说，我感觉我现在在外面没有社恐这一方面的顾虑。在自己的精神世界上不像以前那么悲观了，也不像以前那么抑郁了”
Modified ATs can play a vital role in reducing the social pressure on people with disabilities.

%自信心
\subsubsection{Obtaining Positive and Confident Attitud  Toward Life}
%使用改装后的辅具能给人带来自信。 
Using modified ATs can boost an individual's confidence.
D2 and D3, after using modified ATs, were able to go out, which enhanced their connection with the outside world, their self-confidence, and their love for life. For instance, D3 said his transition from being bedridden three years ago to freely going out with his modified wheelchair (Fig.~\ref{fig:ATgroup} (c)).  ~\begin{quote}
    ~\inlinequote{Three years ago, I was bedridden \joeyrevised{all day,} but after modifying my wheelchair, I could independently go out and enjoy the flowers, and the fresh air. It has improved my self-confidence, making me want to keep living,} he said.
~\end{quote}
%D3-我觉得改变很大，我三年前受伤的时候，基本上都在床上，24小时躺着，三年之后买了轮椅，改装了之后就可以自己独立出去看看路边的花花草草呼吸自由的空气。家里的空气跟外面的空气是不一样的，外面的空气明显就是更甜更更自由，风压力，你窗户关着的话，甚至窗户打开的风，他的循环都不一样。所以我刚才之前提了提高人的自信心，然后而且会改变对生活的态度，就可以接去外面你可以看看烟火气，然后会更想继续活着下去。

The process of modification can also give the person with a disability a positive attitude.
For example, D5 felt a sense of achievement when wrapping his crutch, he had a feeling of a positive and uplifting spirit. ~\inlinequote{Using my white cane makes me feel empowered. Now, I do not see it as cold iron structures but as a source of positive energy for an active life,}he explained.  \joeyrevised{After seeing the completed modifications, D5's parents applaud of his approach and advised him to never give up once he has a goal, and to do everything possible to achieve it.}
%D5-我觉得用起来感给我感觉到一种出去行走什么，就感觉到一种力量或隐形的力量，就会觉得就觉得自己用的这个东西，然后就会感觉到不一样，很不一般这样子，因为毕竟自己做的。然后其次我觉得在使用上我觉得和刚刚买的那个话感觉到一点还是会就不是会感觉到像他我就不会把它感想成那种冷的一个铁矿，铁支架这样子，我会把它想象成就是自己的一个往上积极生活的一个源泉，我觉得是这样子。

%改装这个的话有感而起，想着看了铁的那种颜色也没有，然后就突发奇想，我就想自己在家也闲着无聊，特别那种也比较感觉到有点自己自己无所事事，然后又不怎么想玩手机，然后又不想做什么事情，然后就想着家里面有一些材料，然后就自己DIY这样子。

%S5 also mentioned that the process of modifying devices brought him self-confidence. He stated: ~\inlinequote{Being able to solve the difficulties I face by myself increases my confidence... Sometimes, I even feel I live a more exciting life than able-bodied people.}" 
%S5-他说：“起码自己能动手解决自己所遇到的困难，应该是一个比较自信的过程，对自己的自信心会有一个。”
%S5-从开始彷徨、无助、彷徨，现在以我个人来说，应该无助还是有的，自信心还是有的。就是说在平凡的生活中有那么一点自信，毕竟甚至我觉得我自己有些时候活得比健全人还精彩，对不对？

Additionally, aesthetically pleasing AT modification also helps people with disabilities overcome psychological problems, gaining stronger emotional value and confidence. Before the modification, D1 was very self-abased, \joeyrevised{\inlinequote{I am scared to look other people's eyes, especially when our gaze meet, and I do not let others see me for too long,}} she mentioned.

Now, as a video creator and \joeyrevised{prosthetic eye mutual aid group organizer,} D1 can boldly show off the beautiful prosthetic eye she made in front of the camera (see Fig.~\ref{fig:SNS} (c))  D1 mentioned: ~\inlinequote{My creation of prosthetic eyes is primarily about beauty and confidence, not utility like traditional ATs (see Fig.~\ref{fig:ATgroup} (b)).} 
%She noted: ~\inlinequote{I used to be very introverted due to psychological or mental health issues, hardly socializing, and could not look people in the eye or let them see me for long.} However, the prosthetic eye modification helped her solve these issues. 

%D1-因为我以前可能是在医院关太久了，然后有一些心理疾病或者精神疾病，所以我其实不怎么很少社交。之前就是会非常自卑，然后跟别人讲话都不敢看别人的眼睛，或者不敢让别人长时间看到我。
%D1说-“心路历程一开始是情绪疾病比较严重，因为主要是自卑，然后到后面现在好了很多了，都没有说太过于自卑了，然后做这个事情会心情好一点这样子。主要还是情绪价值。
%D1-其实我制作的义眼它主要还是以美观和增添自信为主，如果是真的辅助设备的话，它实际上是不会往这个方向去研发的，它主要还是以实用性为主。

The use of modified ATs, as well as being in the modification process, can give people with disabilities self-confidence and a positive mindset.

%与社会的联系增强
\subsubsection{Strengthen Connections with the Community by Sharing Modification Experiences}
%大部分改装完AT的残疾人喜欢将自身改装经历分享给所处的社群，从而增强其与残疾人群体的联系。而且这种AT改装经历通常被残疾人群体认为是积极的，从而起到启发其他残疾人或改善其生活态度的作用。
%Most people with disabilities who have modified their ATs like to share their modification experience with their community and enjoy positive feedback, these sharings are often perceived as positive by the disability community and. thereby strengthening their connection with the group. 
%改装完AT的残疾人会将自己的改装经历分享给自己所在的群体，目的是为了帮助到别人，呼吁自己所在群体的残疾人能够变得积极。这些残疾人受到影响并认可，他们从消极情绪中走出来的后也开始进行传播。这样的方式增加了残疾人之间的联系

People with disabilities who have modified their ATs often share their modification experiences to assist others and encourage other disabled people in their group to adopt a more positive attitude. These individuals with disabilities are influenced and recognized and start to spread them. \joeyrevised{This Sharing strengthens the bonds within the community, helping to build a more interconnected network of support for people with disabilities.}

Both D2 and D3 shared their modification results in their communities to help and encourage other people with disabilities. Other people with disabilities followed the method that they shared and got their own modification successfully. D1 shared some videos of herself wearing modified prosthetic eye, urging people in her group not to feel inferior (see Fig.~\ref{fig:SNS} (c)). \joeyrevised{There are some people join her online aid mutual community, she provide prosthetic eye fitting information to others with no charge. %However, the inclusion of some non-disabled individuals with ulterior motives can disrupt the community's atmosphere, so she decided to increase the strictness of the screening process.
}

\joeyrevised{D3 actively shares and improves modification methods within the community, not only posting his own modification videos but also enhancing those of others, thus fostering support among those interested in modifications. As community members watch and share these videos, more people are drawn to join, which in turn provides D3 with recognition and motivation. D3 said: ~\inlinequote{more and more people with disabilities recognize and appreciate this thing, and they would share it with others they know who are in similar situations as mine. By learning from our attitudes and experiences, more people gradually join our group (see Fig.~\ref{fig:SNS} (e)).} Additionally, D3 earnestly responds to inquiries about crucial modification details. This collaborative environment promotes a supportive and positive feedback loop, enhancing the community's collective capacity to adapt and innovate ATs.}

Sharing modified AT experiences strengthens community bonds among people with disabilities, creating a more confident and supportive community where members help each other.

%把分享相关挪前面
%self-fillment
\subsubsection{Finding Vocational Interests and Achieving Self-development}
%知识增长、职业发展
%删掉分享

%不论是残疾人还是其他改装者在改装的经历中，他们的知识都得到了增长。他们有的利用自己的背景知识，成为一个专业改装者，有的甚至通过改装发现了自己潜在兴趣，帮助了自己的职业发展。
Throughout the process of modification, individuals with disabilities as well as other participants engaged in modification activities have experienced a substantial enhancement in their knowledge base. \joeyrevised{Some modifiers take on different roles after their modifications.} Moreover, there are cases where individuals have uncovered previously unrecognized interests during the modification process.

%残疾人的知识在改装过程中得到增长。
The knowledge of people with disabilities has increased during the modification process. For D1, she looked for well-known sources, like information or literature about eyes. In this process, she expanded her knowledge through self-learning and exploration in various ways. When S4 independently completed English Learning Software development, he acquired extensive computer-related knowledge and conducted numerous user needs surveys.
%改装信息渠道其实这个是我完全自发去研究的，我会去找一些比较知名的，就是像北京同仁医院他的那些关于眼睛的那些信息或者文献什么的东西，获取渠道其实挺多的，搜一下信息都有。在我做义眼的机构，它就有很多国外的义眼师，然后我会留有联系方式什么的，然后我就会去咨询。但其实就是咨询我没有说特别多。
%He said ~\begin{quote}~\inlinequote{My gains are increasing, and this motivates me to learn new technologies. If I am unsure about how to implement a certain function, I read about the methods to achieve it, which has also enhanced my programming skills.}\end{quote} 
%Modification enhanced S5's rehabilitation knowledge base as he attended a lot of trainings, which broadened his knowledge.
%S4-收获的话就感觉是跟用户的接触越来越多了，然后也会督促自己会去学一些新的技术，或者说不止就手上所学的东西，可能满足不了这样的一些需求，或者说不知道某个功能是怎么实现的，就要去读一些类似的功能的这种实现的源代码，对编程能力可能是一种促进。 我获得的收获越来越多了，也会督促自己会去学习一些新的技术，如果不知道某个功能如何实现的，我就要去读一些实现这个功能的元代，我的编程能力也得到了促进。 
%S5-他提到：“当时就参加了很多这样的培训，然后自己的知识面也丰富了一点，然后通过自己所学所知再去帮助身边的一些商友，因为帮助的人多了，所以说人家就是我们这个地区的认识我的人也多了，就像小杨介绍你认识我一样，他也是知道这一地区就是说想知道这些内容只有找我，我知道是最全面的，所以他就把我介绍给你了，懂我这个意思吗
%有的改装者有技术相关的背景知识，他们应用自己的知识和能力成为一个专业的改装者。
\joeyrevised{Some, through continuous learning, achieve growth. For example, S5 became a learning delegate and traveled to other cities to attend nursing training classes with the goal of acquiring nursing knowledge to then teach other disabled individuals. He also invited experts to jointly explain nursing knowledge. He organized life rebuilding activities sponsored by ATs companies and was invited to participate in ATs company product promotion activities. There, he incorporated the needs and experiences of actual users into the development and improvement of products. \inlinequote{At that time, I spent five days involved in every aspect from the concepts of the products, through the manufacturing process, sales, to the introduction of their features.} He said.}

%Similarly, S10 shares his production process with those who seek help from him. He mentioned: ~\inlinequote{In fact, a lot of people come over, and if I can make these accessories, they are definitely all willing to pay for them.}
%D3-我会电焊和相关的这些操作，我有20年的经历，我自己也开过店。后来我就优化了一下对方他改了轮椅的方法，后来我在群里要告诉人家，然后我们就会告诉他关键点在哪些地方，应该怎么改，根据你的身高，如果你想躺着，位置就不一样了，你就需要调节。
%S10-其实很多人过来，你要说我有能做出来这些配件，他们肯定是都愿意付钱的。然后我的一般的处理方式是，第一，如果说是我买的，我会把相关链接推荐给他们，如果是我自己做的，我就会把整个流程整个需要用它互做给他说过去。” 
%职业发展
Furthermore, the modification uncovered the hidden interests of people with disabilities, who even found new career directions through the renovation. S4's interest in technology began in junior school when he found computers fascinating. 
He explained: ~\inlinequote{I follow many projects because I am an open-source enthusiast. I enjoy browsing GitHub, where I can also learn various projects.} Drawing from his experiences, S4 joined a high-tech company and now works as a software engineer.
The modification process has helped individuals with disabilities to uncover hidden interests, apply and gain more knowledge throughout the process, and find career paths that suit their development.

In conclusion, modification AT not only enhances comfort and fosters independent living, lightening caregivers' load and bettering family dynamics. It also cultivates a positive, confident mindset, encourages sharing experiences with the community, and aids in discovering vocational interests for self-development.

\subsection{Challenges in Modifying AT}
In the process of searching for more appropriate ATs through interactions with the caregiver, modifier, and surroundings, due to their characteristics, our participants encountered multiple challenges throughout the different stages of modification. These challenges will mainly emerge in three different phases: before the modification, during the modification, and after the modification.

\subsubsection{Before the Modification}
There are three main challenges before the modification: awareness gap, family resistance, and organizational limits.

The first challenge that arises in the pre-modification stage is that people with disabilities are not aware of modifications, the existence of modification products, or basic modification knowledge. Lack of awareness of proper modifications and knowledge of products on the market that can solve the problem at hand can lead to no or mis-opening of the modification process and waste of resources. According to the statement of D7, many people with lower limb disabilities walk for years using benches and wooden sticks due to a lack of information and awareness of available modification aids.
%D7 就是说现在人他有的残疾人他不懂得改装，他根本就不会去改装，知道吧？有的人有的残友我刚才也讲了可怜到出行在家里用小板凳在走路，有的人连拐杖都没有，就用一根木头棍子在那边拄着拐杖在走
D4, as a polio patient who has been using a bench to walk for over twenty years, has used only two ATs for walking, crutches, and a bench. He would only move within the confines of his home due to the inconvenience of traveling before being introduced to a handicapped van by others. 
%D4 大概是在几岁好像是10岁左右，这是左右拿个凳子慢慢的撑一下，甚至什么，就可能可以好像勉勉强可以走路的.拐棍就是一个像我我两个脚都有残疾的，用起来不行的，等于是撑不住的，撑不住还怕是因为滑倒，就是这个道理。公路不好，以前我没有买残疾车的时候也不出去的，等于是在家里的，等于是不你怎么出去，不像我们以前都是石子路，高高低低，还像我拿凳子走，肯定是费很费力的，对吧？再一说你出去不安心走出去，你高高低低不安全的，也不出去的，都是经常在家里的不出去的，我买了残疾车我就等于是那可以开始去买就可以出去了，就这样子的。
These people conduct their lives through reduced standards and physical adaptations. The unawareness is partly due to the lack of social promotion, for example, as S1 said, ~\inlinequote{Some people with disabilities need this thing (modification) but he does not even know about it, the promotion is not enough, including how he can do some modification that may be more convenient, a lot of people do not know about it.} He gave an example, 
\begin{quote}
~\inlinequote{Many people take wood to make a simple toilet, but it is not convenient at all. In fact, this product has been on the market for a long time, it is all very simple stuff, and many people who live in the countryside just do not know about it.}
\end{quote}
%S1 包括一些技术改造，政府项目，包括一些现在我们服务的残疾人，也是政府在买购买服务的，很多人现在是整个的一个理念推广不还不足 ，一些残疾人他需要这个东西，他根本就不知道这个东西，然后根本就没地方，不知道这个市场上有这个东西，推广还不够，整个宣传还不够，包括一些改造怎么样改可能更方便。很多人都不知道。
%S1 对我说的人家现在很多人拿着木头做一个简易的马桶，但根本就不方便。其实市场上很多早就有这个产品，这边也就有了很简单的东西，很多乡下人根本就不知道。

 The second challenge in the pre-stage is rooted in the negative attitude of the family (caregiver) towards modification. The relationship with the caregiver is an important prerequisite for the modification because the caregiver's involvement will be needed throughout the modification process and afterward use. Modification cannot start without support. D3 pointed out that the family's disapproval is the biggest difficulty throughout the modification process. ~\inlinequote{At first, my family didn't believe in it, they didn't support it, they didn't want to go through all that trouble, they didn't want to spend the money.} As the founder of a community of paraplegics, he shared many ideas for modifications for patients similar to him, partly not practiced due to family reasons. ~\inlinequote{Some of my friends in the group would find it too much of a hassle and there are family reasons to just stay in bed all year round.} D3 emphasized that the two biggest factors that make modification possible are the initiative of the person with a disability himself and the will of his family and that there is no way to intervene to change the family atmosphere and conditions.
%D3 家里人不配合，刚开始家里人都不相信。就不支持，或者说是不想这么折腾，然后不想花钱，然后认为做家里人的工作。
%D3 刚开始给他看了一下，也这个人可以有用头开轮椅的，我说我一定也能可以，因为我知道我的头是正常的，跟他一样，只有头是正常的，然后他那个点就没什么障碍，障碍是没改过，但是我是干这行的，是有把握，有不断的家里人心情好的时候就跟家里人做工作。
%D3 他们有家庭条件的，不是说钱的家庭条件，家庭氛围条件的，他们能想改变自己，想改变的人一般都会都会去改变。但是真正人家不愿意嫌麻烦，不想麻烦家里人的，我们也不好直接，因为你改变不了家里人太他态度和他他和他家里的态度就没必要他愿意，这样的话也是他的自由对吧？
%D3 群里面大家都是每个个体，然后生存的每个环境又不同，虽然身体大多数基本上相同，段位已经到了天花板级别，但是所处的环境不同。不是每个人都能坐轮椅出去的，有些人过得也比较苦，说实在的一个人一个故事。

The third challenge comes from the government, an official aid and information access channel, which does not have the means to meet the AT needs and customization demands of people with disabilities. S7, Director of the Adaptation Center of the Federation, clearly stated that the Federation is not able to meet the high standard of customization of persons with disabilities and does not have this service. What the Federation can provide are basic types, of appliances from the list of AT decided by the provincial government. This was also emphasized by D8,~\inlinequote{When I ask about getting a specific white cane at the AT center, people often say they don't know about it; they're more familiar with crutches. After several instances like this, it feels like the ATs relevant to me are mostly unknown to others.} 
%S7 它不属于基本型辅具，我们只能补基本型，所以这种残疾人相对比较高的辅具或者相关的要求，我们就没法满足，这是一个困难，而且没办法解决，因为省市都定得很死，你的辅具配发跟补助必须在基本型辅具目录内，超出目录的一律不行，或者讲你当地残联可以研究一定的操作目录的品种或者什么东西，但我们也不具备这个能力
%D8 像也有一个让我觉得让我觉得很蠢的一个点，想想让我之前有时候说想去辅具中心或者怎么样去买一个盲杖的话，然后经常好像是他会变成是让人赶来赶去的那种，然后一一问到马上人家说不知道，他就说你说啥忙啥，我不知道有些东西存在，我就只知道有拐杖，可能拐杖的话在人们的视野中出现的频率估计是会高很多，所以这样的话这样的次数多了点以后，然后我就感觉好像是跟我有关的辅具，人们都说不知道，可能就慢慢的也就没有想就没怎么想起说去了解这方面的事情
The government, as a more prestigious and enforceable platform, cannot meet the demand for customization, which means that people with disabilities who seek customization need to complete the whole process of personalized modification by themselves and without supervision and support from official and credible channels, which will be a long and difficult process.

\subsubsection{During the Modification}
There are three main challenges during the modification process: people with disabilities physical conditions limit modifications, difficulty in finding the right assistant, and limited resources and craftsmanship. 

The first challenge in the modification is mainly the inability of the participants to continue modification on their own due to the presence of their disabilities. First, when the participants are proposing the modification program, their physical limitations can lead to an inability to further express the design and requirements through means other than words, ~\inlinequote{Before the modification, I conceptualized it in my brain, but when I asked someone else they didn't know how to do it, and I could not move my hands myself to make a sketch (D2) (Fig.~\ref{fig:ATgroup} (a)).} Second, when the participants are the operators of the modification program, their physical limitations make them unable to complete the process independently and require additional help from others. D5 illustrated that he had his parents involved throughout the modification process, and due to his lower limb disability, he was less efficient or unable to complete the process of his operation. A similar situation occurred with D2's modifications, where his father helped him with a series of modifications including assembly (see Fig.~\ref{fig:ATgroup} (a)). Both of the participants needed to secure another alternative solution or support from others, to make sure that the whole modification process went smoothly.
%D2 在大脑里面在自己的脑海里面模拟了数遍以后，感觉可以了，定型下来，因为找别人别人也不知道怎么做，然后毕竟我自己手又动不了，画不了图纸
%D5 这时候父母也帮着弄，我母亲帮我弄的居多一些，就提供一些材料给我，然后我没有对照他模子来，我就是按照自己想的就自己包自己弄这样子。那个时候大概弄了一两个小时，因为自己脚也不方便，有可能弄的时间比较长，
%D2 老爸帮我组装了，因为你去这就是做一些也就是找电焊师傅看好的一些东西回来，然后在轮椅上面打孔，用固定螺栓固定好，就是这样。现在自己动不了，没办法，只能自己提出方案，让别人去替我操作。

The second challenge that arises during the modification process is the difficulty of finding the right individuals to help with modification due to the immature business model. The specific details will be discussed in Section~\ref{BMTSec}. The process of searching for participants with a disability puts them in a vulnerable position, left to passively accept the decisions of others.

%和下面这个结合 提到前面}
The reason for the inefficiency is the scarcity of material resources for the modifier, which makes it difficult to find suitable modifying materials. Firstly, there is limited access to materials. S3 suggested that ~\inlinequote{a lot of the materials are hard to find, and you end up finding other things to replace them. I wanted to use similar ones when I watched some videos online, but I could not find a channel to buy them.} S5 also emphasized that ~\inlinequote{American companies have the best cushion materials, but many domestic manufacturers are unable to make similar products.} In searching for materials, modifiers want to find materials that meet multiple goals at once, but it is very difficult. ~\inlinequote{I hope it can be a little bit lighter, now it is iron as a fixing, which is stronger if it is replaced by aluminum alloy, but I reckon it must be easy to break (D7).} 
%S3 但是因为有些材料确实不好找，最后还是找的其他东西来代替的。本身观看网上的一些视频的时候，想用类似的材料，但是没有办法购买，或者是可能性价比不高，所以之后用了木板去进行自己制作。
%S5 现在目前市场上说坐垫的话最贵的也是最好的，没有企业它的确实是最好，它防压疮效果是最好的，是一个美国产的漯河roho品牌叫roho。他做的坐垫的话是最好的，但是国内现在它是一个叫什么材料的一个叫绿什么橡胶的一个材料，但是国内现在好像好多厂家在仿，都仿不出来，都仿不出来。
%D7 希望它能够再轻一点，因为现在它全是铁，然后铁上面还要用更厚的铁作为拉拉住它，固定它，很牢固，其实他这些我整个的前面没有，后面坐后面太重了，我就希望能不能用什么铝合金材质，什么不锈钢材质，但是我估计肯定它容易弯，它容易折断
%S10 首先轮椅材质它这个东西估计都没法焊，再一个也不知道用它就很多，你看有些需要加弯管的，有些需要加角的东西，咱估计得到专业的地方去做了。
%D7 我要花1000块钱再换一个锂电池，我拿不出这么多的钱知道吧？（D7）——该参与者无工作
%S8 我感觉对于未来可能就是在材料上，我希望能有更多的材料能更适用于咱们所有的这些项目，因为现在比如说咱们想实现一个东西，然后去找材料的时候，可能都得按照现有的这些材料，如果说它材料在材料这一方面能你比如说咱们想用钛合金，但是钛合金的价格可能太贵了，所以能不能就是说对于材料这方面有一个需求，还有一个说咱们整个国家算是可能考虑对残疾人再多一点关爱，感觉跟他们接触的多了，可能比较共性，感觉还是对他们的关爱就是少一点。
A significant factor contributing to the low efficiency of modifications is the limited craftsmanship of the modifiers, with only a small portion of the envisioned modifications being feasible. S10 stated having considered welding during the modification process but lacking expertise in that area, resorting to using needle and thread or ropes for fastening. Additionally, he expressed a desire to adjust the support for the arms on the wheelchair, but the complexity and difficulty of the required operations made it unattainable. S5 mentioned that certain operations requiring special tools are beyond his capability, leading to a current feeling of helplessness.

%S10 然后焊接在一起就没法折叠了，所以说需要焊接一种那种类似于插管的方式，在两个把手这放两个圆筒，然后再插进去，这么一个高靠背，然后随时取下来，中间也其实想过自己焊，但是确实这方面不太懂，因为我经常也看一些抖音，就是那些手工活打神，他们做东西做的可好，但是确实没有这方面天赋，我们后来都是有时候都用针线或者用绳子捆绑的方式去做。
%S10 整个也是就在一个微调整，其实我觉得还能改得更好，但只是确实脑子里没这么多东西，比如说它架胳轮椅架胳膊的地方，我看安维轮椅的，他们加胳膊地方也能调整，随时调整，往往里收往外推往哪走，但对我们来说改装太困难了，基本上实现不了。
%S5 比如说你轮椅的话，当时设计的时候前后座高不太好，你是不是需要焊接？对不对？比如说一个轴承锈死了，锈在里面了，你取不出来，那东西可能需要一个什么轴承拉子什么把它拉出来，这些东西我们手上没有的话，也是很无助的，搞不出来。

The third challenge is that modifiers have a limited knowledge base related to modification. D1 emphasized that ~\inlinequote{the material technology (for prosthetic eye modification) is very complicated and it takes a lot of lessons to understand how to do this stuff} and~\inlinequote{I'm not a professional either, it takes a lot of chemical knowledge and it is just a matter of trial and error.} D7 also pointed out how during his modification, he had to change the wheels frequently in the beginning to continue using them because he did not understand the difference in wheel materials. This all goes to show that modifications require specialized training and a knowledge base. Limited knowledge related to modification can lead to difficulties in choosing and executing design solutions. S5 suggested that without professional knowledge of welding, it can cause safety hazards, so it is not recommended to carry out this aspect of retrofitting on top of a wheelchair. Without experience and design background in modifications for people with disabilities, it is difficult to produce products with high application value. S4 mentioned that the difficulty in the whole process is that none of their team has a professional background in designing for people with disabilities. Finally, even though some modifiers search for information through the Internet, it is still scarce, and D3 mentioned how the Internet can be considered a blank canvas for information on modifying specific wheelchairs.

%D1 就是材质技术这些东西。如果你想知道的话，你其实可以对义眼机构了解一下，然后或者说他们也有培训班什么的，你可以自己看一下，因为它这个很复杂，就很可能要上很多节课，你才能明白这个东西怎么做。（D1-304）困难还是挺多的，主要还是刚刚提到过材质，它有些材质没有办法跟一眼很好的结合，然后这个也是只能靠不断的试错，因为我也不是专业的，他需要很多化学方面的知识，当然我不是很擅长。（D1-321)
%D7 一开始我不懂，就是一开始的这个是我的第三次换轮子的知道吧？一开始的时候换的都是PVC的，他那个有的时候还能用我两三年，然后前两年买的 PVC的用了一年，基本上就磨平了，没用。后来在某宝上面我就挨家挨户的在看，在选，然后就问人咨询人家在淘宝上对淘宝这样居安之的，他人家滑冰的滑轮，你应该见过滑轮的轮子的材质，它就是聚氨酯的，它防滑能力也比较强，你在反角度的情况下，你推他的时候，他是不会那么的轻松的能推得动他。以前我还不懂这个。(D7-40-46)
%S5 专业知识的缺乏影响设计方案的选择。我觉得那样子改装大的改装以后反而对它轮椅结构，如果说你对一个轮椅的结构进行改装以后，反而你没有你没有正规的一些知识点，比如说电焊对吧？比如说铝焊，你到时候焊焊不好，你反而会给他产生安全隐患，说不定在路上滚滚轮子掉了，对不对？也是有，我们也遇过这种情况的，滚一滚滚滚大梁断了，对吧？所以说我倒是不建议在轮椅上面，在轮椅车上，我觉得大费周折我觉得没有意义的。(S5-141)
%S4 困难的话缺钱其实对我觉得还是一个可能这方面的我们都不是很专业，没有太多这方面专业的人，我们这个是这样的同学，然后有工程师，但是都是比较缺乏这方面设计的这种背景，专业的这种专业性可能不是特别强。
%S6 人体工程学，人机交互。对，它其实它不是一个单学科的问题，它是一个像国外有好多学者提这种超学科的概念，比如说交叉学科，但是社区他们是因为他没有一个好的平台，他们都自己搞，他们没有一个平台，他就没有钱，没有流动资金，没有他们的这种规模性不是特别大，起到的作用也不是很大，他们他们那些装置都是机械结构或者一些小的东西。
%D1 最大的困难和挑战就是它不稳定性，我不确定我要做多少颗，才能做到一个完全适配，给他的一眼也不确定他想达到一个什么样的效果。因为每一颗都是手工的，它不是一个完流水线的那种生产，没有标准。(D1-444)
%D3 从网上面找相关面的资料，但是网上面对这方面的资料可以说是一片空白，基本上没有。(D3-118)

\subsubsection{After the Modification}
There were three main challenges in the use phase after the modification: physiological discomfort in use, more trouble for family members with the modified aids, and difficulty in adapting the aids to complex external environments. Ultimately there will be two ways to deal with these challenges: performing repeated iterations of the modification and compromising.

The first challenge after the modification is the physical discomfort of the user during use. During the use phase after completion of the modification, ~\inlinequote{discomfort} appeared frequently in the descriptions of S10, D2, and D6. ~\inlinequote{We wanted to add a high backrest and made a plate directly on the backrest of the wheelchair, she was uncomfortable sitting ...... then just discarded it (S10).} D6 also noted that ~\inlinequote{it is just uncomfortable, it is too big, and sometimes it often runs out on its own like that.} This bad experience is often blamed on a lack of modification skills and associated knowledge base. Of these, the lack of knowledge about the progression of the condition needs to be emphasized. S10 stated that there is a correlation between cushion replacement and the progression of buttock muscle atrophy in people with disabilities. ~\inlinequote{Generally neck strength also decreases as the condition progresses and then we need to add a high backrest (S10) (Fig.~\ref{fig:ATgroup} (f)).}
%S10 比如说第一代的时候，我们想的是增加一个高靠背，最我记得是直接做了一个板子，然后直接放在他轮椅靠背的，他坐着不舒服，然后我们之前是没有，包括他现在高靠背是确实买了一个带有高靠背的轮椅，它上面有个这么一插这样一个东西，之前没有这个东西，然后后来但是轮椅我们坐着不舒服，然后推着也不舒服，因为我需要轮椅是这种一体式的，比较轻便的，我可以我经常拉着我自己拉着轮椅就能拉着我妈上下楼梯去过任何的路，但那种轮椅就不行，那种轮椅一压起来可能要散架或者是不太好。
%D6 其实说实话不太好，因为我感觉还没有找到适应我的，之前我爸可能是自己做的，然后两个相结合的，试用了一段时间还不错，但是就是不舒服，也就是太大了，有时候撑的鞋子有时候经常自己跑出来这种。
%S10 它一般他脖子力度也会随着病情发展减弱，然后我们给他增加了一个高靠背，是买了另一个轮椅上一个高靠背，然后放到另一个低靠背的轮椅，它有个轮椅坐的舒服，靠背上还放了一个枕头。

The second challenge in post-modification is that using modified ATs can cause endless hassles for the family. D3 noted that ~\inlinequote{you have to set up the phone stand to play with the phone and charge it from time to time (Fig.~\ref{fig:ATgroup} (c)), and thirdly you have to ask others to pick up the stick that you have in your mouth(to play your phone), for example, if you drop it... These are details that may seem trivial, but they are very important for the family to understand.} Especially in the use of the wheelchair, ~\inlinequote{my type of pathology will cramp, and people will slowly slip down. The distance between the lever of the controller and the chin would change, and the family would get annoyed if they adjusted it every day.} Additionally, the use of ATs and modifications to the environment will also bring about a change in the environment for the family and they will need to make adaptations. 
~\inlinequote{For example, many of our indoor bathrooms nowadays have ramps added for the convenience of people with disabilities, but the family will find it inconvenient to make a ramp at home...... There is also the aesthetic problem of adding a handrail in the dining room, which the family will find too ugly and doesn't fit in with the decoration.}

%D3 甚至有些玩不了手机的，他能玩手机，但是他家里人不支持他玩手机，因为这你玩手机你必须要摆摆好手机支架对吧？第一点手机支架你要摆好，第二点你要时不时还要帮你充电，然后第三点你含在嘴巴里这根棍子，你必须比如掉了，你要帮他捡一下，帮他拿一下掉到床上，对，你够不着的位置叫家人帮忙打。这些小细节，小事情，看似小事情，但是对于我们这种患者就比较了解。
%D3 我没有第一代第二代、第三代，我就一个轮椅，然后就改装了一次，只是刚开始改装调试好了我就开，我觉得可以开动了就ok了。后来时间长了就发现，因为我这种病理型的它会抽筋，会人会慢慢往下溜，往往下抽筋之后就要往下坐，控制器的操纵杆的话，它就会位置就慢慢跟跟下巴的距离就会改变。后来就叫家里人调适左右，上下，要选一个固定的位置，大概中间的位置，偏中间一点的位置，就不用每天去调的那种位置，这样的话就更好。因为每天去调，家里人也会烦躁。
%S1 比方说我们现在很多卫生间，比方说它是住在室内的，它可能会有个高差，我们觉得他要做个波段给他做好了，他们觉得家里面做个波段不方便，就给他敲掉，是这样，碰到很多的。还有一个美观度，比方说我装一个扶手，餐厅是确实是需要的，但是家人觉得比如说大厅里什么装一个扶手就难看了，现在因为乡下很多地方，但是装修的还可以的，你装了一个扶手，他觉得跟他的装修风格不搭，或者每天美观什么的，各种问题是很多的。因为残疾人他也需要一个家庭环境，他也不是说每个残疾人都是个体生活的。

the third challenge is that because the use of ATs often occurs in multiple scenarios, it has difficulty adapting to complex external environments resulting in inoperability. D6 noted that the modified product was able to be used in the summer, but would be inoperative in the winter. D2 emphasized that as soon as a first modification was made to go out and use it in a bumpy situation, the fixture would loosen up and fall apart and become inoperative. This was particularly the case in his environment where the park was older and the roads were broken. S10 thought ~\inlinequote{I need a lighter wheelchair because I often pull my mom up and down the stairs to any road by myself.}
%D6 可能是想了，因为现在的话差不多都是裸露表面，因为你像现在夏季对吧？其实还好，但是你如果说像以后像冬天什么的话，就可能说之前就是用的布，但是缠着的话就不跟你说很难受这种感觉。然后所以说想着能拿像硅胶或者说这种软塑料，然后去就形成一个这种在做一个这种比较能够适用于鞋里面的这种可以说去户外可以使用的这种
%D2 然后我就按照他那个视频，从网上面买配件，就是买夹具钢管之类的回来改的，就像我们刚开始的说的时候，用一我自己没找别人用电焊焊的时候，只是买家具买钢管这些改装的，可以在家里面使用，但是不能出门，一出门的时候他就一颠簸肯定会出现松动的现象，就散架了。
%D4 公路不好，以前我没有买残疾车的时候也不出去的，等于是在家里的，等于是不你怎么出去，不像我们以前都是石子路，高高低低，还像我拿凳子走，肯定是很费力的，对吧？再一说你出去不安心走出去，你高高低低不安全的，也不出去的，都是经常在家里的不出去的。
%S10 因为我需要轮椅是这种一体式的，比较轻便的，我可以我经常拉着我自己拉着轮椅就能拉着我妈上下楼梯去过任何的路，但那种轮椅就不行，那种轮椅一压起来可能要散架或者是不太好。

As mentioned in Section~\ref{testSec}, almost all assistive devices require testing and continuous iteration. The negative experience of testing modified AT ultimately led to two different responses: repeated iterations of modified products or compromises on all fronts.D2's three generations of iterations for wheelchair cushions are a prime example of the former response (see Fig.~\ref{fig:ATgroup} (a)). The first generation of cushions was used by sewing a cloth cover on the outside of a found sponge, but it slipped easily due to a poor fit, which led to the second generation of cushion modifications, which was a solution to the non-fitting problem by searching for an anti-pressure sore cushion on the internet. But time will be deformed. The third generation of the cushion is replaced with a combination of a car cushion and an anti-pressure sore cushion. The effect is better. The second way of coping is to compromise, usually by reducing one's experience of use to accommodate the aids. For example, D3 suggested that ~\inlinequote{people need to adapt to the wheelchair, although some of the features of the wheelchair do not fit and it cannot be helped if they cannot add them.} D7, on the other hand, was still confined to sitting on a reclining bar when using the adapted ATs due to the unavailability of specialized ATs on the market that met the requirements and the inappropriate sitting height (see Fig.~\ref{fig:ATgroup} (d)).
%D2 第一代：家里面自己做的搞海绵。从从小区里有很多装修的拆下来的海面，就是自己捡回来，然后缝合一个布套。但他有一样不好就是它容易滑，因为它贴合性不好。第二代：就从网上搜索的关键词，比如说防压窗坐垫它就是有那种坐垫，前面高后面低的那种，正好和整个人的屁股比较贴合的那种形状的——但是那种做一做它就是变形了.第三代：最后就是直接换成汽车坐垫，从拆旧厂找的汽车坐垫换在上面，然后上面再加一个防压窗的气垫，就感觉还挺好的。
%D3 因为你主要还是人要去适应轮椅，虽然轮椅一些功能不符合，但是你要看能添加就添加，不能添加。你像我岳父他们的老年人，他也去他会去适应轮椅，而不是轮椅来适应他。
%D7 降低自己的使用体验去迎合辅具 - 市面上没有符合的专门的辅具，坐下高度不合适，只能坐在斜杠上。
\subsection{Business Model for Modifying AT}~\label{BMSec}
As previously mentioned, the Chinese AT modification market faces issues such as low awareness among people with disabilities, poor purchasing power, and small market sizes. Therefore, professional AT modification companies are still very scarce in China. As S2 stated:
\begin{quote}
\inlinequote{There indeed exists a significant gap between the current market and the emergence of companies specializing in AT modifications, and this may be a problem that cannot be resolved in the short term.}
\end{quote}
 %S2-我认同你这种说法，它（当前市场离专门做辅具改装的公司的出现之间）确实是一个鸿沟，而且可能短时间内还解决不了的问题。
However, there still are some pioneers. They spontaneously explored business models related to AT modification based on their respective situations. This section will introduce five categories of AT modification institutions (see Table.~\ref{BMtab} in the Appendix) to gain a better understanding of China's AT modification market.

\subsubsection{Standard AT Manufacturers with Modification Services} AT manufacturers typically have well-established production lines, making them generally well-equipped for modifications, some of them also have engineering departments, allowing them to fulfill personalized AT customization requests. However, in many cases, manufacturers may choose not to establish independent modification businesses due to economic, safety, and cost considerations. As S2 mentioned, ~\inlinequote{It (the AT modification industry) does not have such a high-profit margin, so there are not many people willing to do this.}
%S2-因为如果说你要从比如说残障人士出行的角度去看这个课题的话，我觉得应该是比如说因为你看一定是像这些事情，因为像轮椅的出席一定是有轮椅公司去提供这种这种服务的，它不会有轮椅在改装公司的，没有这种公司，它没有像汽车改装汽车行业这么大，也没有这么高的一个溢价，所以不会有人来做这件事情，能会做这件事情的人和公司只有电动轮椅公司手动轮椅公司以及辅具公司，不会有其他公司来做这个事情的。
Therefore, currently, people with disabilities mainly rely on after-sales services provided by AT manufacturers to achieve their modification goals.

HUWEISHEN~\footnote{\url{https://mall.jd.com/advance_search-1804804-10513383-10303719-5-0-0-1-1-24.html?other=}} is an electric wheelchair manufacturing and selling company that operates in this manner. As the boss of HUWEISHEN, S2 told us: ~\inlinequote{Out of the 100 units we normally sell, there might be 8 or 10 with this kind of demand (modification requirements), or even more.}
%S2-很多是比如说在我们正常销售100台里面，可能就有10台或者是8台，再有这种需求，甚至是会更高一些。
According to S2, most of the modification demands for electric wheelchairs involve common personalized requests, such as adding a commode, widening the seat, or attaching a trailer, among others. These types of modification requests only require simple processes, and therefore will be addressed directly by HUWEISHEN's production workshop. 

Furthermore, within HUWEISHEN's wheelchair donation program, they also modify wheelchairs for recipients based on their individual needs. S2 mentioned:~\inlinequote{Our goal set for our company is to donate 50 electric wheelchairs in a year... We may prefer to donate to younger people or those who need electric wheelchairs more.}
%S2-我们给自己公司的定下的目标是一年的话是要捐赠50台电动轮椅出去的，这个是针对不同的，我们可能更愿意去，比如说圈到偏更年轻的那种，更需要这种电动轮椅的人去。
D9, a man with limited upper limb mobility, is one of the beneficiaries, and HUWEISHEN helped him modify his manual electric wheelchair into a toe-operated electric wheelchair. These highly personalized and more complex wheelchair modifications are handled by the engineering department, which communicates with people with disabilities to design suitable wheelchairs and then produces them in the production workshop. However, S2 also acknowledged, \inlinequote{Cases like D9 (complex modifications) are rare... We try to meet the needs that users present, but if the research and development cycle is too long, we may not proceed.}

\jtrevised{From our participant's report, we can see such large-scale standard AT manufacturers have stable revenues and specialized AT modification technology. However, although they have significant advantages in providing personalized modifications of ATs, their ability to offer widespread customized services is limited by operational costs. Moreover, standard AT manufacturers themselves are quite rare, typically emerging only in markets with a substantial demand for assistive technologies.}
%S2-这种很特殊，这种情况很少...当然有一些还是一个有些用户他给我们提了需求，我们也是尽可能的去满足他，只要这个需求我们能做得出来，但如果说涉及到研发以及周期很长的话[xO7] ，这个事情我们可能就做不了，比如因为大部分刻制化的东西，可能也就涉及到比如说一些焊接，一些泽湾，一些其他的比较简单的刻制，我们都是不需要花太多的一些成本的。

\subsubsection{Unregistered NGOs} 
Due to the difficulty of registering NGOs, there are a large number of unregistered NGOs in China~\cite{zhang2017nothing}. It is more difficult for them to obtain social donations and government funds. Yangzhou Hope House, founded by S5, is one of them. Yangzhou Hope House is an organization composed of a team of 60 volunteers. The main business of this organization includes two aspects: First, there is a consultation and shopping service. The organization selects suitable ATs or accessories for customers based on their needs. Second, there is a modification service. The organization replaces and assembles ATs or accessories for customers' existing equipment based on the purchased items. %People with disabilities as customers only need to pay for the cost of the ATs or accessories themselves.

According to S5, the main income of this organization comes from the owner's monthly work-related injury compensation and another unrelated business (selling tobacco and alcohol). In the case of AT modification, their modifications are done on a charitable basis. People with disabilities only need to purchase the required additional parts and do not have to pay for the modifications.

S5 told us that this organization has unique business advantages: ~\inlinequote{As fellow sufferers (people with disabilities), they (other people with disabilities as customers) are introduced by the disability community to us, and then we offer them a lower price than Taobao, so they are willing to buy our service. In addition, the upstream manufacturers around us are also willing to sell products to us because we have after-sales service (referring to free modification services).}
%S5-现在就是说我们作为一个商友，就是商友圈就是说这样子一个介绍，然后我们身边因为他考虑到价格方面，因为我们给他们价格肯定是比淘宝上还要便宜，所以说他们也考虑找我们，我们给它的价格又便宜，售后服务又有保障，所以说一般的我们身边的上游都是找我们。
%However, according to S5, in addition to the profit from selling standard ATs or accessories, he also relies on the profit from selling tobacco and alcohol and his monthly work-related injury compensation to subsidize this Hope House organization.

This organization's modification mainly revolves around wheelchairs, including replacing the front and rear wheels of electric wheelchairs, backrests, and brake pads, among other things. S5 also acknowledges that their organization has limited AT modification capabilities. He said: ~\inlinequote{We only carry out optimizations, such as replacing your wheelchair's backrest or the wheelchair's battery. We are limited to modifications in this regard.} 

%\hkrevised{Organizations founded by individuals with disabilities, who have similar experiences, often understand the needs of people with disabilities better. Additionally, they usually manage their own communities and keep in communication with them, giving them a more complete set of experiences and information to accurately determine the real needs of people with disabilities. Furthermore, due to the lower cost of parts and free retrofitting services, individuals with disabilities also enjoy reduced modification costs. However, these organizations face limitations in their capabilities, such as a lack of professional engineering and medical personnel, as well as the absence of a stable business model to help a wider range of people with disabilities.}
%S5-不是我们自己做的，也就是我们把一些实用的东西拿回来之后，给它进行一个优化，进行一个优化。比如说靠背，你说靠背你自己改装，你说我们自己买材料回来做吗？不现实的，我们买了靠背回来，比如说旧的东西我们可以改装，比如说你靠背你做了三年以后，你里面的海绵什么的，不好了，我们可以把套子拆掉对吧？给你里面的海绵可以把它更换一下，或者说电池里面电芯不行了，我们可以把电芯更换一下，懂吗？是只限于这方面的去改装。S5-我觉得那样子改装大的改装以后反而对它轮椅结构，如果说你对一个轮椅的结构进行改装以后，反而你没有你没有正规的一些知识点，比如说电焊对吧？比如说铝焊，你到时候焊焊不好，你反而会给他产生安全隐患，说不定在路上滚滚轮子掉了，对不对？也是有，我们也遇过这种情况的，滚一滚滚滚大梁断了，对吧？所以说我倒是不建议在轮椅上面，在轮椅车上，我觉得大费周折我觉得没有意义的。

\jtrevised{According to our participants, such unregistered NGOs are a historical product of Chinese NGO policy. They are primarily mutual aid communities founded by people with disabilities themselves. Individuals with experience in modifications altruistically provide relevant information and assistance. However, the lack of policy support, sustainable profit models, and professional rehabilitation technology support constrain the sustainable development of these organizations.}

\subsubsection{Registered NGOs}\label{Sec:Registered_NGOs}
Due to the unique development path of the disability sector in China, the growth of government-independent registered NGOs providing charitable services for people with disabilities is currently in its nascent stage. 

In this context, S7 established the Makingforgood Community~\footnote{\url{https://www.makingforgood.cn/gywm}}, an NGO that established a sustainable business model to provide free AT design and modification services for people with disabilities. S7 stated:
\begin{quote}
~\inlinequote{Our organization is the first and only of its kind in China. We produce personalized ATs for free for people with disabilities. There were similar organizations before, but they shut down because they couldn’t sustain their business model.}
\end{quote}
%S7-所以说像我们这种机构在中国我们是第一家，目前也是唯一一家，我们给残疾人免费制作辅具，之前也有类似的，但是都关掉了，它形成不了 sustainable。
According to S7, their innovative business model can be summarized as a combination of free modification and paid education. The organization has a total of 20 paid staff, including 5 full-time employees and 15 part-time employees, and a peripheral volunteer team of about 1,500 people. The 15 part-time employees mainly participate in the production phase of AT modification, and they all have a good knowledge background like Tsinghua University and work experience like IBM China, Xiaomi, etc. Furthermore, the organizational structure of the institution consists of an executive leadership team, an education team, and several project teams. The executive team is responsible for assessing and analyzing the needs of people with disabilities, forming project teams based on volunteers' skills, and the educational team transforms the AT modification projects into teaching projects to develop participants' research, design, prototyping abilities, and specific skills like programming. The participants' tuition fees support the operation of the entire organization.

Regarding the current business status, as per S7's interview and the website information, the Makingforgood Community's design and modification projects include prosthetic limbs for individuals with missing hands that can operate computers and smartphones, gait trainers for cerebral palsy patients, and custom leg supports for an individual with a lower limb amputation, D10. These project processes resemble the Double Diamond model in design methodology\cite{council2005double}. They often start with need assessment and analysis by the executive leadership team, followed by solution design, team formation based on the solution, entering the prototype phase, prototype testing, and finally delivering the iterated final product to the clients. S8, a long-term volunteer (with some remuneration) in the community, summarized: ~\inlinequote{We customize things from scratch, and also modify existing ATs.} 
%S8-因为咱们从算是0从0~1去制作一个东西，所以说在技术上肯定会遇到一些小问题，但都是小问题，因为咱们大方向只要定好了之后，这个东西确定能实现了，剩下的都是一些个小问题。也有改装，有从0~1的制作也有改装。
S7 added:~\inlinequote{We manage about 50 AT modification projects a year. We already have people with disabilities waiting in line for our modifications, but we lack the time and resources.}
%S7-啊，不做推广，就是我们满足不了那么多的需求啊，就比如说我们现在每年能做五十个。啊，那么满足五十个呃辅具的需求，就这个五十个残疾人的需求。那其实我们现在已经有排队的了，我们是时间精力不太够。
Additionally, S7 shared their project standards: ~\inlinequote{Safety comes first. Will the products we make be dangerous? How dangerous? Effectiveness comes second. We don’t want our products to be black boxes, like for autistic children. Currently, there is no consensus internationally on the causes and treatment methods for autism.} Additionally, according to S7, their projects aim to enhance the original functionality or experience based on safety and effectiveness. He mentioned: ~\inlinequote{We made a chopping aid for upper limb amputees. Since they only have one arm, they can't control the ingredients and have to use their feet, which can lead to falls. Our aid helps them control the ingredients, thereby increasing safety.}

%\hkrevised{Makingforgood, a registered NGO, boasts a well-developed professional team and workflows for research, design, and production, offering free customized design and retrofitting services to people with disabilities. This attracts many clients, but due to limited service capacity, they provide services selectively. They typically create prototypes of assistive devices using traditional manual techniques or simple welding, prioritizing high safety standards. According to source S7, they rarely customize devices for individuals with lower limb disabilities due to the significant risks involved, and they do not initiate projects with uncertain outcomes. Furthermore, since the organization sustains itself by converting assistive device projects into educational initiatives, projects with potential educational outcomes are given precedence.}

\jtrevised{From our participants' reports, we can see that these emerging registered NGOs are gradually rising. For example, Makingforgood, an NGO founded by scholars combining education and DIY-AT community operations, has experimented with some potential sustainable business models in the Chinese AT modification market. However, such philanthropic organizations are still in their nascent stages in China, and currently operate on a relatively small scale.}
%S7-安全性第一个。就是他他做的要求，我们做这个东西会不会有有危险性？对危险性有多高？这是主要的。嗯。那么。呃，第二个就是有效性。我们不希望做做黑箱的东西。就是大家做了很长时间，然后也投入很多很多时间和精力，发现这个东西一点用没有。啊，那所谓黑箱呢，我们现在。呃，不碰的，就是跟可能跟自闭症啊相关的。嗯，也有人在就是呃找我们想做这种自闭症呃沟通的，帮助他儿童沟通的，呃，提高他社交能力的呃这种小的玩具。但是自闭症。呃，就从从成因到治疗手段，在国际上都没有一个定论。
%S7-你如果不给我做的话，我就现在就这么对付我，我还是拿个拐杖，我没有腿托，我就直接把腿放到那个拐杖的那个扶手上。我还是会这么做。啊，因为我不可能晚上上个厕所，我要把那个假肢穿上啊，所以。i have nothing to lose.嗯，哎。嗯，那么他现在就是用这个东西啊。他可能是。危险性是百分之五十。啊，那我我给他做了一个东西。安全性可能提升百分之。十百分之二十对他来说也是很好的。包括舒适性啊。啊，他会要求百分之百的解决。所以我们做产品数据定制。这个就是呃，不是。跟这种上市产品。一个理念。那这个东西要去上市，你你你需要。可靠美观。干嘛？对了，哥，那我这个东西其实就是解决它最根本的一个东西。嗯，就像我们给残疾人做这种。切菜上肢残疾。切菜的辅助器。那他只有一个胳膊的时候。他就没有办法辅助食材，有的时候他用脚去放到菜板上去按着。他的食材按一份肉，按着菜。啊，按一个西红柿，然后再去切。啊，那么他一个脚站在那就有可能滑倒，他有可能把脚切到。对吧？那我们给他做一个东西，如果我让你不用一个脚在上面去按着。啊，你用另一种方式。那我这个东西。就是我不用做的那么精美。啊，我只要不给你造成伤害那

\subsubsection{Independent Software Development Group} Information and Communication Technology (ICT) should be regarded as a fundamental human right for people with disabilities, as an integral part of the lives of the majority of the population \cite{borg2011right}. In this study, two different software applications were developed to address the specific needs of visually impaired individuals.

S6, a visually impaired individual, participated as a user tester in the development of a virtual white cane application. S6 explained the principle of this app: ~\inlinequote{It utilizes the laser radar sensor of an Apple to assess the surroundings, then employs various sensory cues like sound and vibration to alert users about the distance from obstacles, creating a meaningful mapping.} According to S6, at the time of the interview, it had already been adopted by thousands of users. To maintain the normal use of the APP, they have charged fees, but whether it can operate normally remains to be verified. S6 also pointed out the challenge of obtaining valuable feedback to iterate the software when visually impaired individuals serve as users:
\begin{quote}
~\inlinequote{Some of the data from visually impaired users may not be accurate... We receive relatively limited feedback that truly provides valuable information.}
\end{quote}

%\hkrevised{To sum, for independent developers, gathering high-quality feedback from the disability community in the field of digital assistive devices is increasingly challenging. Development teams for these devices must create additional channels to reach users and engage in multiple rounds of communication to collect more accurate feedback.}

\jtrevised{We can observe that with the rise of ICT technologies, individuals with disabilities who are interested in software development can conveniently create customized solutions for themselves and others with similar needs. However, software development is systemic engineering work, and individual developer groups also face challenges in sustainable operations.}

\subsubsection{Independent Technicians with Modification Skills}~\label{BMTSec}
Independently technicians are an important segment of the AT modification market. They usually have another identity in life and have certain hands-on abilities, such as maintenance, welding, etc. However, the business model of independent technicians is even more immature. 
%D2-轮椅我基本上是改过几次，我来看一下。第一次的时候我就没有想到找电焊师傅，第一次的时候我就是从网上面买了各种各样的管夹，夹具，自己回来，然后用钢管自来水管，用夹具各方面拼起来的，当时用也可以，只能在家里面，比如说路面一颠簸的可能就不行了，毕竟那些夹具是夹不住的，最后我才想到找电焊师傅，按照我的方法去焊，焊一个就是固定控制器的支架，包括家具，这样改装以后效果就挺好。当时电焊师傅只是就是焊了一个固定控制器的支架的夹具，但是这个夹具看上去是非常牢固的，每次拧紧了以后看上去非常牢固的，但是你在外面行驶的时候，电动轮椅在外面走的时候，本身我像我这样的腰背肌没有力量坐在轮椅上面，就是东倒西歪的那一种。
Difficulty in pricing is the first problem. \inlinequote{The complexity of the craftsmanship of customization is far more complex than normal craftsmanship, and it does not make sense for people with disabilities to charge high craftsmanship fees, especially the short life span of the use of the customized AT (D1).} For the technicians, D2 mentioned that:\inlinequote{My technicians told me that the modification is time-consuming, but because it is a modification for a person with a disability, he is embarrassed to charge more (see Fig.~\ref{fig:ATgroup} (a))}. As mentioned above, the stereotype is the second problem. Due to stereotyping, D2 was rejected by many independent technicians because they thought that this kind of contact with people with disabilities was best avoided. After all, it was embarrassing, and even if they did it, they did not want to communicate with people with disabilities in person. 
Lack of extensive experience in AT modification is the third problem. This results in uncertainty about the consequences of modifications, as they may fear taking responsibility and therefore decline modification requests. D3 indicates that the technical personnel mentioned they had not encountered this before and could not handle it well, politely refusing. Similarly, in S5, there’s a related situation: the person who repairs electric bikes is unwilling to fix the wheelchair because they’re unfamiliar with it. 

\jtrevised{
From the reports of the participants, it is evident that independent technicians with modification skills are crucial for AT modifications. They possess the modification skills needed by people with disabilities, but yet they are the group most lacking in a business model to drive their service provision. Luckily, as mentioned in Section~\ref{Sec:Registered_NGOs}, emerging NGOs are in fact attempting to connect these individuals and are developing payment methods to motivate them to perform modifications.
}

In conclusion, through a specialized case study on the business models of AT modification, we can see that the five business models currently existing in the Chinese AT modification market are all quite preliminary. As shown in Table.~\ref{BMtab} in the Appendix, these five types of organizations each have their advantages and disadvantages. Commercially-driven AT manufacturers, despite having strong modification capabilities, often exhibit a low willingness to engage in modifications, resulting in a limited scale of modification operations. In the case of NGOs, whether registered or not, their modification capabilities vary depending on the circumstances, but all require a sustainable business model to support the expenditure of AT modification services. For independent software development groups, although there are fewer restrictions on software development, obtaining high-quality feedback for software iteration poses a new challenge. Furthermore, whether their software revenue can sustain development and operational costs requires further research. As for independent technicians, although they are abundant in daily life, they often have limited involvement in AT modification due to biases and a lack of modification experience.

\section{Discussion}%kexin
In our findings, we articulate the entire process of modifying ATs in China, starting from the motivation---such as the need for customization due to limited modification options and personal interests---to the detailed process involving collaboration and potential conflicts of interest. We explore the psychological benefits of these modifications and the challenges faced, notably the reluctance of skilled people under social pressures. Furthermore, we explore the business model where manufacturers profit from standard devices to fund limited customizations. 
These findings comprehensively address the research questions we posed. Most importantly, these findings not only align with previous research but also unveil new insights specific to the Chinese context, leading us to propose several crucial points and their design implications.

\subsection{The Importance of Personalized Manufacturing of ATs}
%motivation+benifits@ kexin
%summary相关的finding
%findings和related work的关系，哪些是一致的。
%哪些是我们unique的发现，这些发现的重要性是什么？2~3段解决

Our research highlights the importance of personalized modifications in ATs addressing specific needs. For instance, individuals D6 and S10 were able to adapt their standard devices to a usable and more comfortable level through modifications. This finding is consistent with previous studies. Given the diverse environments and evolving needs of people with disabilities, continuous customization of ATs is crucial to reduce abandonment rates, as indicated in existing literature~\cite{hocking1999function}.

Our study further underscores the significance of modifications, especially under constrained economic conditions. Case D7 illustrates this: abandonment of an AT equates to a loss of independence, yet prolonged use often leads to decreased stability and other functional issues. In such scenarios, modification becomes the only viable option, a common predicament in low- and middle-income countries with significant wealth disparities.

A novel aspect of our findings is the emphasis on the psychological benefits and new identities fostered through the act of modification, a factor previously overlooked in research. While past studies have focused on designing ATs for social inclusivity and self-esteem of people with disabilities~\cite{li2021choose}, our research reveals that modification is more than a means to an end; it is a form of self-expression for people with disabilities. Participants enjoyed the process of modification, feeling empowered, optimistic, and confident. This transformation was often attributed to the modified device itself, as expressed by D5:~\inlinequote{I envision it as a wellspring for my upward and positive journey in life.} Furthermore, the success of these modifications inspired them to spread their techniques and positive mindset, transitioning from being just users of AT to becoming innovators like screen reader modifiers (S4), prosthetic eye designers and influencers (D1), and professional device fitters (S5). This shift underlines the empowering process of involving people with disabilities in the design process, affirming their ideas and capabilities, and enabling them to participate in societal matters related to disabilities actively, thus amplifying their influence.

This study further explores the positive impact of assistive technology on collaboration and interaction among individuals advocated by CSCW. Previous research in Human-Computer Interaction has indicated that the involvement of individuals with disabilities, family members, friends, and caregivers in the design and implementation of DIY-AT can improve relationships~\cite{hook2014study,hurst2011empowering}. Similarly, in our research, D3 highlighted that the successful implementation of DIY-AT mitigated conflicts with caregiving family members arising from ongoing trivial matters. By gaining more independence through the use of ATs, individuals could reduce dependency on caregivers and achieve more harmonious intimate relationships. This aligns with previous research and underscores the specific relationship between DIY-AT and collaboration. Additionally, our study adds a new perspective by emphasizing the importance of the process itself for disabled individuals collaborating with others in DIY-AT endeavors. Previous research has cited uncertainty about the outcome of modifications as a major obstacle to DIY-AT~\cite{slegers2020makers, hook2014study}. In our research, S10, as the implementer of wheelchair modifications for his disabled mother, found great happiness in interacting with the user, discovering issues together, and discussing specific DIY solutions during the process. Compared to the outcome, the greater freedom of the DIY-AT process allows for more discussion and communication to occur.

\subsection{Business Model for Personalized Manufacturing of ATs in the Absence of Philanthropic Cultures}
%summary相关的finding @Haokun
%findings和related work的关系，哪些是一致的。
%哪些是我们unique的发现，这些发现的重要性是什么？
%这里可以参考第一篇推荐文献
The unique contribution of this study is its bottom-up portrayal of the \hkrevised{business model of} AT modification sector for people with disabilities in China, especially against the backdrop of the commonly held belief that the development of services for people with disabilities in China is challenging~\cite{kohrman2005bodies,stone1996law, vaughan1993development}. \hkrevised{Past research has shown a lack of sustainable business models in the AT service sector, attributed to reasons such as limited market size, lack of user-centered design, and policy implementation issues~\cite{oderanti2016holistic}. In China, issues with policy implementation and lack of focus on end-users also persist. Additionally, although China's vast population of people with disabilities presents a huge potential market, the AT industry still faces challenges due to the unique characteristics of the Chinese market. }

\hkrevised{
Firstly, in terms of funding, people with disabilities are generally economically disadvantaged as consumers, and this holds true in China as well. } This further supports the view that ATs have a social welfare aspect and are difficult to develop through market mechanisms due to pricing issues~\cite{jiang2023development}. Also, Chinese NGOs as suppliers find it even more challenging to obtain government and social assistance. Due to issues such as an immature charity culture and limited government funding~\cite{zhang2017nothing,weller2004civil,wang2001development,guo2022remote}, their sources of income are very limited, making the development of their own business models a necessity. The case of unregistered NGOs, like Hope House, also sheds light on the difficulties of NGO registration in China, resonating with the findings of previous research in this area~\cite{zhang2017nothing}. \hkrevised{Furthermore, the Chinese AT industry primarily consists of low-end standardized products that are unevenly distributed, with varying standards~\cite{wang2023discussion}, which also affects the willingness of people with disabilities to consume.  Against this backdrop, there is a general lack of research on the business models of AT service organizations in China, and our study has documented and reported on the commercial situation of various organizations in this field, particularly providing a detailed description of the business situation of Chinese NGOs, thereby filling this gap.}

\hkrevised{Secondly, the commercial sustainability and business risks of AT services are also concerns of the CSCW community. One study in the community investigated the manufacturing, importing, and distribution of assistive devices in Bangladesh, revealing significant challenges in the commercial sustainability of the AT industry in the absence of an official distribution system and financial support~\cite{khan2017assistive}. Also, Siny Joseph et al. attempted to establish an economic value assessment tool for AT services from a cost-effectiveness perspective~\cite{joseph2020assessment}. Tigmanshu Bhatnagar et al. found a partner-centric business model in their research on the AT market in Africa. Startups adopting this model created a sustainable and scalable business model for their products~\cite{bhatnagar2023s}.  Our research uniquely contributes by categorizing and organizing the business models of AT service organizations in China, particularly NGOs, filling a gap in the landscape of this area in China.}

Finally, the NGO Makingforgood Community in our study also created a viable business model in AT modification by developing it into an educational project for students through a teaching team to cover the cost of AT modification service. \hkrevised{The model of achieving sustainable business value by establishing a broader stakeholder network for collaboration (incorporating teaching teams and students into the business model) also contributes to the CSCW community.}

%本研究的独特贡献在于它自下而上地描绘了中国残疾人辅助技术（AT）改造部门的情况，尤其是在普遍认为中国残疾人服务发展具有挑战性[39, 60, 64]的背景下。这些组织独立开发了各种商业模式来启动AT改造业务。尽管如此，个性化的AT改造部门仍面临诸如商业成熟度低、规模小、残疾人购买力低等问题。这进一步支持了AT具有社会福利性质的观点，并且由于定价问题，很难通过市场机制发展[35]。
%此外，本研究的独特贡献在于它考察了中国参与残疾人AT改造的非政府组织（NGOs）的当前生存状况，从而推进了对中国致力于残疾人事业的NGOs的研究。两个这样的NGOs自给自足的需求表明了中国NGOs在确保资金方面的挑战，无论是来自政府还是社会捐赠[66, 68, 72]。像希望之家这样的未注册NGO的案例，也揭示了中国NGO注册的困难，这与该领域先前研究的发现相呼应[72]。
%最后，一个可行的商业模式对于产品或服务的可持续性至关重要[14, 57]。由NGO Appro TEC与设计公司IDEO合作为肯尼亚小规模农民设计的水泵的成功，取决于创建了一个为网络内多个利益相关者增加价值的商业平台。这个平台协助甚至培训了新的参与者，如当地分销商和修理技术人员[14]。X也强调了多利益相关者网络不仅有助于创建和部署DIY-ATs，而且还有助于确保这些技术的长期成功和可持续性。同样，在这个研究中，NGO Makingforgood社区也通过将AT改造发展为教学团队的学生教育项目，找到了AT改造的新价值，以覆盖AT改造服务的成本，最终创造了一个可行的商业模式。

\subsection{Design Implications}
%4-5句，最好能举一些可行的例子（引用）

%残疾人康复知识融入设计，junyi
\subsubsection{Integration of Rehabilitation Knowledge in AT Design}
In our study, despite some people with disabilities lacking rehabilitation knowledge, there is a clear demand for such knowledge among them. 
%For example, S5 proactively sought advice from experts to learn rehabilitation knowledge, which was then shared within the community with other people with disabilities.
\joeyrevised{Due to the physical changes experienced by the mothers of S10, D5, and D2, among others, there is a need for extensive iteration of their assistive devices. For example, D2 and S10 underwent multiple modifications to address issues with wheelchair cushions and high backrests.  Throughout the modification process, they had to consider various factors, such as the impact of time on the components, environmental effects on fixation methods, and especially the impact of physical changes on the assistive devices. In wheelchair design, we discovered significant deficiencies in standard wheelchairs in meeting individualized needs. The insufficient flexibility in wheelchair components prevented adaptation to changes in user physique. Due to a lack of design and engineering background, adjustments were only reactively made when problems became apparent. This highlights a design challenge: how to create flexible, adaptable assistive devices that can adjust to changes in users' physical conditions while also considering environmental and temporal factors. To address the creation of flexible, adaptable assistive devices, we propose that combining rehabilitation knowledge with design may allow for predictive adjustments and support a more proactive design approach. By integrating this knowledge early in the design stage, necessary adjustments can be considered, future iterations predicted, and this forward-looking design strategy can effectively reduce the costs and inconveniences of subsequent modifications, thereby enhancing product usability and user satisfaction.}
%设计师考虑护理知识在设计过程中的发现阶段能够起到更多维度的信息

%involve co-design 残疾人的协同决策者的重要地位,junyi
\subsubsection{Caregivers Participate in Co-design as Stakeholders}
Previous research has shown that incorporating people with disabilities in the co-design process is crucial for creating ATs~\cite{bircanin2021including,wang2023can}, however, few have mentioned the importance of including caregivers as co-decision makers.

In our study, S1 mentioned that the aesthetic preferences of family members were affected by certain modifications for accessibility, leading to reluctance in cooperation and necessitating changes to the modification plans. This experience indicates that this is a common issue, highlighting the importance of considering caregivers' needs in the modification process.

Participants like D9, D7, D5, D3, D2, and D1 require a familial environment, not just an individual living space. \joeyrevised{They cannot live apart from their family and need to share the same environment with them. Some modifications to ATs can impact the family members.} Additionally, the compromises made by family members for people with disabilities should not be taken for granted, recognizing that each person in the household has their own needs. \joeyrevised{S1 mentioned that during the process of making his home more accessible, ramps were built to eliminate level differences between spaces to facilitate movement for disabled individuals indoors. However, these ramps were eventually removed because other family members felt that they interfered with the convenience of daily life. }

Therefore, involving caregivers as stakeholders in the co-design process is equally important. \joeyrevised{Including caregivers as co-design stakeholders not only enriches the design process but also ensures that the resulting ATs are more aligned with the actual needs and environments of the end-users.} \joeyrevised{A participatory design approach could be considered, where all family members, including caregivers, are given a voice in the design process. This approach helps identify and harmonize the diverse needs and preferences of all household members, leading to more universally acceptable solutions. Co-design sessions could include discussions on how ATs affect family dynamics. Visions might involve designs that facilitate interaction, such as ATs provide communal spaces that are accessible and comfortable for all users, including non-disabled people. } Accordingly, it is recommended that designers consider the involvement of caregivers as relevant stakeholders in collaborative design.

%规范标准化辅具方便改装配件的生-kexin
\subsubsection{Standardizing AT Accessories to Simplified Customization} The diversity of user groups and the breadth of assistive products present a complex issue in the provision of AT. Ensuring that as many people as possible have access to assistive products and services, enabling them to better engage in society and interpersonal interactions, is a highly significant matter. Universal Design, with its focus on creating products and environments that are inclusive of diverse populations, can be introduced into more AT design and development processes~\cite{hitchcock2003assistive}. This approach not only enhances the usability and effectiveness of assistive devices but also promotes social inclusion and equality for all users~\cite{rose2005assistive}.

In low- to middle-income countries, where economic resources are scarce and there is a shortage of knowledgeable care providers, initiating AT from design schemes poses a significant challenge for a large portion of the population~\cite{de2018assistive}. Therefore, for disabled individuals lacking specialized knowledge, a feasible and cost-effective solution is to meet their specific needs through the purchase of readily available components. Applying universal design while producing the ATs is a possible way.

Participant S2 highlighted a key factor inhibiting AT manufacturers from expanding into customization services: the lack of standardization in the AT market, particularly wheelchairs. The market is flooded with numerous brands, each adhering to different standards, making it impossible to use uniform parts for repairs across different brands. This lack of standardization creates technical and component barriers, leading to high costs in customization. When a customization request is made, manufacturers often have to start from scratch in parts development.

To mitigate this issue, standardizing AT accessories and unifying accessory specifications, such as the size of screws, could facilitate widespread application of modification parts. This approach would distribute the high costs of customization across multiple factories, rather than concentrating them in one. Consequently, manufacturers with the capability to customize would be able to expand their service offerings. For buyers, this means having the option to select and customize ATs that better meet their individual needs at a similar price point to standard models. This would represent a significant step towards more accessible, personalized AT.

%政府加强商业模式的引导建立宣传-okun
%\subsubsection{A viable business model matters}
%商业模式的重要性，有商业变现模式，才能有更多NGO企业诞生，
%特别是在中国，强政府，NGO受政府的引导，政府应该帮助相关企业建立商业模式。

\subsection{Limitation and Future Work}
%被试的范围和数量，未必能generalized
%？

In this study, we explored the processes and methods of DIY-AT among people with disabilities in China. While our participants were distributed across various regions in China, their overall number was relatively small and thus may not adequately represent the entire population of people with disabilities in the country. Moreover, the majority of our participants had incomes below the national average, which aligns with previous research suggesting that DIY-AT can fulfill unique needs at a lower cost. However, understanding the perspectives and decision-making processes of people with disabilities from different economic backgrounds could provide new insights and dimensions to this area.

Additionally, our recruitment of participants, particularly those serving as "suppliers" of modified equipment, was limited to male providers. This gender-specific characteristic of our sample might impact the comprehensiveness and impartiality of our findings. Furthermore, our study did not include all categories of disabilities, particularly individuals with cognitive impairments. Future research should aim to broaden the scope to encompass a wider range of people with disabilities, thereby gaining a deeper understanding of the intricacies and challenges of DIY-AT for different groups.

Our study indicates an in-depth analysis of the interactions between social organizations, government, people with disabilities, caregivers, and other stakeholders about customized AT. Given the scarcity of research on DIY-AT in developing countries, especially China, our work contributes new discourse and insights in this field. Future research could build upon our study by further refining the needs of different types of people with disabilities, leading to more precise recommendations. This approach will not only enrich the existing literature but also potentially inform policy and practice in the development of AT.

\section{Conclusions}
This study adopts a qualitative research methodology, utilizing semi-structured interviews conducted in China with 10 suppliers and 10 demanders in the context of DIY-AT for individuals with disabilities. Our findings suggest that in addition to motivations like comfort and reducing caregiver workload, personal hobbies and social media also drive AT modifications among people with disabilities in China. During the modification process, most individuals collaborate with family members and skilled technicians. We utilized an adapted double diamond model to summarize their modification and iterative processes. Additionally, we have summarized the benefits and challenges associated with AT modifications. We found that modifying AT not only enhances comfort and mobility for people with disabilities but also fosters improvements in their psychological experiences and self-fulfillment. These insights are crucial for understanding the significance of AT modifications and inspiring future AT design practices. However, during the modification process, they face a range of challenges related to awareness, knowledge, family, and social support. Specifically, addressing the sustainability challenges of AT business models, we conducted a specialized case study on the organizational forms of modification services identified in our study. Our research provides valuable empirical evidence on the current use of AT among disabled individuals in China, which can inspire future universal AT designs.

%%
%% The acknowledgments section is defined using the "acks" environment
%% (and NOT an unnumbered section). This ensures the proper
%% identification of the section in the article metadata, and the
%% consistent spelling of the heading.
\begin{acks}
This project is supported by the National Natural Science Foundation Youth Fund 62202267. 
We thank the non-governmental organization \textit{Makingforgood Community}~\footnote{\url{https://www.makingforgood.cn/gywm}} for generously providing the resources. Sincere thanks to all our participants.
\end{acks}

%%
%% The next two lines define the bibliography style to be used, and
%% the bibliography file.
\bibliographystyle{ACM-Reference-Format}
\bibliography{sample-authordraft}

%%
%% If your work has an appendix, this is the place to put it.
\appendix

\section{Demographic Information of Participants}
% Please add the following required packages to your document preamble:
% \usepackage{booktabs}
\begin{landscape}
\begin{table}[!h]\small
\renewcommand{\arraystretch}{1}
\caption{Demographic Information about Demanders}
\resizebox{20cm}{!}{
\begin{tabular}{m{1.25cm}<{\centering}|m{0.5cm}<{\centering}|m{0.75cm}<{\centering}|m{2.25cm}<{\centering}|m{2.25cm}<{\centering}|m{1.5cm}<{\centering}|m{4cm}<{\centering}|m{1.5cm}<{\centering}|m{1.5cm}<{\centering}|m{4cm}<{\centering}}
\toprule

Participant & Age & {\makecell[c]{Gender}} & Highest Level of Education             & Occupation                                                       & Family Monthly Income  & Disability                                                                                          & Congenital or Acquired & Types of ATs Involved in Modification    & Specific Modification Actions                                                                                                             \\ \midrule
D1          & 29                      & F      & Bachelor's degree                      & independent media blogger                                        & prefer not to say      & Removed the right eyeball, there is no sense of distance when looking at things                     & Congenital             & artificial eye                           & prefer not to say                                                                                                                         \\ \midrule
D2          & 40                      & M      & high school/technical secondary school & unemployed                                                       & below 450 US dollars   & High-level cervical spinal cord injury, now only the head can move throughout the entire body       & Acquired               & electric wheelchair                      & Modified the electric wheelchair controller to be positioned near his head                                                                \\ \midrule
D3          & 43                      & M      & high school/technical secondary school & worker                                                           & below 450 US dollars   & High-level quadriplegia, unable to move limbs. Loss of control over bowel and bladder functions     & Acquired               & electric wheelchair                      & Positioned the controller near the chin; installed a phone holder, a stylus holder, a mobile panoramic camera as a dashcam and wheelchair cushion               \\ \midrule
D4          & 59                      & M      & junior high school and below           & worker                                                           & below 450 US dollars   & Childhood paralysis, no sensation below the waist, unable to walk                                   & Acquired               & wooden bench                             & Customized the length, height, and width of the wooden stool, as well as the angle of the legs and adding non-slip features to the bottom \\ \midrule
D5          & 22                      & M      & high school/technical secondary school & worker                                                           & 750 to 1200 US dollars & Leg injury, unable to walk independently                                                            & Acquired               & white cane, wheelchair                   & Decorated the cane by wrapping it with yarn, covered it with old clothing, and adding stickers for beautification                         \\ \midrule
D6          & 22                      & M      & high school/technical secondary school & company employees                                                & 750 to 1200 US dollars & Toe deformity with the right big toe and middle toe pointing upwards, unable to walk or run quickly & Congenital             & toe straightener                         & Made rubber toe correctors by using molds                                                                                                 \\ \midrule
D7          & 30                      & M      & junior high school and below           & unemployed                                                       & below 450 US dollars   & Osteogenesis imperfecta, prone to fractures, incomplete bone development, short stature             & Congenital             & toy car for children; electric wheelchair & Replaced the wheels and frame, and added a basket to the car; replaced the battery in the wheelchair                                      \\ \midrule
D8          & 30                      & F      & high school/technical secondary school & massage therapist                                                & prefer not to say      & Visual impairment with some degree of blurred vision                                                & Acquired               & reading stand                            & Modified the lid of a cup with a cover to use it as a reading stand                                                                       \\ \midrule
D9          & 28                      & M      & junior high school and below           & unemployed                                                       & 450 to 750 US dollars  & Severe cerebral palsy, level one disability, completely unable to take care of daily life           & Congenital             & electric wheelchair                      & Modified the electric wheelchair to be foot-controlled                                                                                    \\ \midrule
D10         & 50                      & M      & Bachelor's degree                      & Chairperson of the Association for People with Limb Disabilities & prefer not to say      & left leg amputated from the hip                                                                     & Acquired               & white cane                               & Added a shelf to the crutch to support the buttocks                                                                                       \\ \bottomrule
\end{tabular}
}
\end{table}
\end{landscape}

% Please add the following required packages to your document preamble:
% \usepackage{booktabs}

\begin{landscape}
\begin{table}[!h]\small
\renewcommand{\arraystretch}{1}
\caption{Demographic Information about Suppliers}
\resizebox{19cm}{!}{
\begin{tabular}{m{1.25cm}<{\centering}|m{0.5cm}<{\centering}|m{0.75cm}<{\centering}|m{2cm}<{\centering}|m{1.5cm}<{\centering}|m{1.5cm}<{\centering}|m{1.5cm}<{\centering}|m{1.5cm}<{\centering}|m{3cm}<{\centering}}
\toprule
Participant & Age & Gender & Highest Level of Education             & Major                           & Occupation                      & Disability Situation & The Roles in AT customization & Specific Modification Actions                                                                                        \\ \midrule
S1          & 34                      & M      & Bachelor's degree                      & Architecture                    & Accessibility Facility Engineer & No                 & Manufacturer                  & Accessibility Environment Retrofit                                                                                   \\ \midrule
S2          & 38                      & M      & high school/technical secondary school & N/A                             & Wheelchair Industry             & No                 & Manufacturer                  & Electric Wheelchair Modification                                                                                     \\ \midrule
S3          & 21                      & M      & Bachelor's degree                      & Electrical Automation           & Electrician                     & No                 & Individual technician         & Wheelchair Modification, Adding Off-Road Wheels                                                                      \\ \midrule
S4          & 24                      & M      & Bachelor's degree                      & Acupuncture and Massage Therapy & Software Engineer               & Visual Impairment  & Individual technician         & Develop English Learning Software for the Visually Impaired, and Enhance the Non-Visual Desktop Access Screen Reader \\ \midrule
S5          & 41                      & M      & high school/technical secondary school & N/A                             & Unemployed                      & Spinal Cord Injury & Individual technician         & Adding a Front Attachment, Modifying the Seat and Installing a Rear Light on the Wheelchair                         \\ \midrule
S6          & 20                      & M      & Bachelor's degree                      & Medicine                        & Student                         & Visual Impairment  & Individual technician         & Suggestions for Optimizing the Infrared Cane App                                                                     \\ \midrule
S7          & 40                      & M      & Bachelor's degree                      & Journalism and Design Studies   & Designer                        & No                 & Nonprofit Organization        & AT Design                                                                                                            \\ \midrule
S8          & 33                      & F      & Bachelor's degree                      & Automation                      & Engineer                        & No                 & Individual technician         & AT Customization                                                                                                     \\ \midrule
S9          & 43                      & M      & Bachelor's degree                      & Services for People with Disabilities    & Civil Servant                   & No                 & Government                    & N/A                                                                                                                  \\ \midrule
S10         & 30                      & M      & Bachelor's degree                      & Software Engineering            & Unemployed                      & No                 & Individual technician         & Wheelchair and Toilet Chair Modification                                                                             \\ \bottomrule
\end{tabular}
}
\end{table}
\end{landscape}

\section{Modification Organization Information}

\begin{landscape}
\begin{table}[!h]\small
\renewcommand{\arraystretch}{1}
\caption{Organization Information and Capability Summary}
\label{BMtab}
\begin{tabular}{m{1.5cm}<{\centering}|m{2.25cm}<{\centering}<{\centering}|m{1.75cm}<{\centering}|m{1.75cm}<{\centering}||m{1.25cm}<{\centering}|m{1.25cm}<{\centering}|m{1.25cm}<{\centering}|m{1.25cm}<{\centering}|m{1.25cm}<{\centering}|m{1.25cm}<{\centering}|m{1.25cm}<{\centering}}
\toprule
{\multirow{2}{*}[-6ex]{Category}}&{\multirow{2}{*}[-6ex]{\makecell[c]{Source of \\ income}}}&{\multirow{2}{*}[-6ex]{Enterprise Info}}&{\multirow{2}{*}[-6ex]{\makecell[c]{Categories of \\ ATs \\  covered}}} & \multicolumn{7}{c}{AT modification capability}                                                                                                                                                                  \\  \cline{5-11} 
{} & {}& {}& {}& Demand Analysis & Software development & Basic AT accessories assembly & Basic processing like welding, cutting & Design \& Customization & Physical prototype production & Factory customization \& Production \\  \midrule 
Standard AT manufacturer                   & Standard electric wheelchair sale                                                                 & Over 40 employees, and an operating income of over 100 million & Electric wheelchair, and related accessories                                 & achieved               &                      & achieved                                           & achieved                                      & achieved                       & achieved                             & achieved                                   \\ \midrule
Unregistered NGOs & Sales such as standard auxiliary tools, tobacco and alcohol, and the founder's work injury salary & Over 60 volunteers                                             & Mainly wheelchair and related accessories                                    & achieved               &                      & achieved                                           &                                        & achieved                       &                               &                                     \\ \midrule
Registered NGOs      & AT modification teaching project sale                                               & 20 employees, and over 1500 volunteers                         & Various types of ATs                                           & achieved               & achieved                    & achieved                                           &                                        & achieved                       & achieved                             &                                     \\ \midrule
Independent software development group              & Software engineer job salary, and English learning APP income for visually impaired people        & Independent developer                                          & The English learning application for visually impaired people                & achieved               & achieved                    &                                             &                                        & achieved                       &                               &                                     \\ \midrule
Independent technicians & Labor wages such as maintenance or welding & Usually self-employed & Depends on demand &   &   & achieved  & achieved  &    &    &   \\ 
\hline
\end{tabular}
\end{table}
\end{landscape}

\section{Interview Guidelines}

\subsection{Interview Guidelines for Individuals with Experience in Modifying ATs for Others}

Hello! Thank you very much for participating in our interview session. My name is xxx, and I am from the xxx. This is our group's first time conducting interviews involving individuals with disabilities, and we are not very familiar with this field, so please forgive any inadvertent offensive remarks. The purpose of this interview is to understand your usage and modification of assistive devices. Some questions during the interview may overlap with the content you have already discussed, and we hope you can understand. We will pay you approximately 8 dollars per hour for participating. The interview will last about fifty minutes. Throughout the interview, if you feel any discomfort, you can skip or terminate any question at any time. Your responses will be kept strictly confidential and used solely for scientific research purposes. May I record this session?

\subsubsection{Basic Information Collection}
\begin{itemize}
    \item Can you tell me your name and age?
    \item What is your current occupation?
    \item Can you tell me about your educational background?
\end{itemize}

\subsubsection{Specific Details About Disabled People}
\begin{itemize}
    
\item When did the symptoms start?
\item How do the symptoms affect daily life?
\item When did you start using related assistive devices?
\item Which assistive devices have you used?
\item How has your experience been with using these devices?
\item When choosing assistive devices, which factors were important in your decision?

\end{itemize}

\subsubsection{Background of the Interviewee}

\begin{itemize}
    \item Do you have a background or interest in mechanics or ergonomics?
\end{itemize}

\subsubsection{Motivation for Modification}
\begin{itemize}
    \item Why did you decide to make these modifications?
    \item What problems did you encounter with the original product (which aspects of the original equipment made the experience poor for disabled users)?
    \item What does modifying a wheelchair mean to you personally? What kind of changes has it brought for disabled persons?
    \item Did you follow any specific guidelines when modifying the wheelchair?
    \item Has anyone expressed surprise or appreciation for your modified wheelchair? If people on social media have inquired, have you offered them modification suggestions or personally modified for them?
    \item As a family member (or someone close), what are the advantages during the entire modification process?
\end{itemize}

\subsubsection{Modification Process}
\begin{itemize}
    \item What preparations were made (did you look for relevant information before modifying)?
\item How did you communicate with the user?
During the modification:
\item What was modified, and which assistive devices were modified?
\item Can you describe the specific operations?
\item What tools were used during the modification process?
\item What difficulties were encountered during the modification, and how were they resolved?
\item What design decisions were made based on? (Your own experience/use, daily observations)
After the modification:
\item What was the usage like for the first generation after the modification? Were you satisfied? Was it as expected?
\item Was there a second generation of modifications, or was it abandoned? Why?
\item What have you gained from the entire process? What aspects would you like to improve to make the modification more perfect?
\end{itemize}

\subsubsection{Psychological Impact of Modification}
\begin{itemize}
    \item What changes has the modification brought about for you?
\item What are your views on modification?
\item What are the social and local community's views on modification?
\end{itemize}

\subsubsection{Vision and Expectations for the Development of Modification Technology}
\begin{itemize}
    \item What are your hopes or expectations for the future development of assistive modification technology?
\end{itemize}

Thank you very much for taking the time to participate in our interview. This concludes our interview. Do you have any other questions?

\subsection{Interview guidelines for Individuals with Disabilities Who Use and Are Involved in the Modification of ATs}

\subsubsection{Basic information collection}
\begin{itemize}
    \item Can you tell me your name and age?
    \item What is your current occupation?
    \item Can you tell me about your educational background?
\end{itemize}

\subsubsection{Symptoms and Life}
\begin{itemize}
    \item Can you briefly describe your symptoms and their progression?
\item How do your symptoms affect aspects of your life?
\end{itemize}

\subsubsection{A Typical Day}
\begin{itemize}
    \item Can you describe a typical day or an event in your daily life?
\item What are your daily activities like, including leisure activities?
\item After the prosthetic eye modification, what daily life processes have changed from being unachievable to achievable?
\item In your daily activities, how do you interact with others?
\item What does a typical day at work look like for you?
\item Can you describe in detail the process of going out?
\end{itemize}

\subsubsection{Living Environment}
\begin{itemize}
    \item Who do you live with?
\item What should be considered in your living environment following your symptoms? What precautions have been taken to prevent collisions?
\item Have there been any adjustments in your living environment?
\item What things do others help you with in your life?
\end{itemize}

\subsubsection{Motivation for Modification}
\begin{itemize}
    \item What problems did you encounter during the use of the original product?
\item  Can you explain in detail why you decided to modify?
\item  What does the modified assistive device mean to you personally?
\item  How has the modified assistive device changed your life?
\item  Did you pay attention to anything particular during the modification?
\item  Has anyone expressed surprise or appreciation for your modification?
\item  How do you evaluate your own modification?
\end{itemize}

\subsubsection{Modification Process}
\begin{itemize}
    \item Where did the inspiration for the modification come from?
\item  What are your channels for modification information?
\item  What problems have you encountered with the specific technical solutions for the modification? Who did you consult?
\item  Through what channels do you seek to advance your technical skills?
\item  Do you regularly learn about assistive devices and emerging technologies?
\item What kind of preparations do you make before starting the modification?
\item Can you describe the specific process of your modification?
\item What difficulties did you encounter during the modification process? How were they resolved?
\item What problems did you encounter with your specific technical solutions? Who did you consult?
\item Through what channels did you seek to develop your technical skills?
\item As your condition progressed, how did you adjust your expectations for the modification?
\item What was the use like for the first generation of your modified device? Are you satisfied?
\item Is the effect of the modification as you anticipated?
\item Have you made a second generation of modifications? What was the reason?
\item After completing the modification, what have \item gained from the entire process (materially or spiritually)?
\item What aspects of the modification do you still wish to improve?
\item What are your visions and expectations for the development of assistive device modifications?
\end{itemize}

\subsubsection{Impact of Modification on Spirit}
\begin{itemize}
    \item How have your symptoms changed from the beginning to now?
\item In this experience, have the goals of using assistive devices and modifications also changed?
\item After experiencing the prosthetic eye modification, what changes have you noticed in yourself? Have your thoughts changed from before?
\item What are the societal and personal perspectives on this?
\item How has this changed your life?
\end{itemize}

\subsubsection{Business Model}
\begin{itemize}
    \item Would you perform similar modifications for others?
\item Do you charge for modifications for others? \item How do you price them?
\item How do your clients find you?
\item Where do you communicate with them?
\item What problems have you encountered when modifying for others? How were they resolved?
\item What are the biggest difficulties and challenges you face when customizing for others?
\item Do you see business opportunities in assistive device modification?
\end{itemize}

\subsubsection{Publishing Video Content}
\begin{itemize}
    \item What motivated you to start publishing video content?
\item How do you feel about the content you publish now compared to before?
\item How do you view comments from online users?
\end{itemize}

\subsubsection{About the Community}
\begin{itemize}
    \item How did you find the community of other modifiers?
\item How do you interact with each other in the community?
\item What are your feelings after joining the community?
\end{itemize}

Thank you very much for taking the time to participate in our interview. This concludes our interview. Do you have any other questions?

\subsection{Interview Guidelines for Representatives from Third-party Modification Agencies and Non-profit Organizations}

\subsubsection{Basic information collection}
\begin{itemize}
    \item Can you tell me your name and age?
    \item What is your occupation and job description at the third-party modification agency?
    \item Can you tell me about your educational background?
\end{itemize}

\subsubsection{Background on Institutional Modifications}
\begin{itemize}
    \item What services and products ar generally provided? Is there a focus on any particular service?
\item Who are the services aimed at?
\item What motivated the start of assistive device modifications?
\item Could you describe the founding and development process?
\item Could you describe the organizational structure, such as first-line and second-line responsibilities? What makes the recruited employees different?
\end{itemize}

\subsubsection{Specific Modification Process}
\begin{itemize}
    \item How do you find effective clients?
\item Through what channels do clients typically contact the institution?
\item How are specific modification needs confirmed during the early stages of modification? How is communication handled, and who is responsible for it?
\item What does the specific modification process look like? What difficulties have been encountered? 
\item Which departments are involved in the collaborative effort during modifications?
\item How is the usability and scientific basis of the design ensured?
\item What does the post-sale process look like?
\item Could you explain the entire business model? What are the standards for fees and the post-sale process?
\item How does the specificity of custom modifications affect the entire transaction process?
\end{itemize}

\subsubsection{Support Related to Modifications}
\begin{itemize}
    \item  Is there support from the government and related agencies?
\item How is support provided?
\item What industry standards must be adhered to?
\end{itemize}

\subsubsection{Visions and Expectations for the Development of Modification Technology}
\begin{itemize}
    \item Do you usually keep up with other assistive technology developments?
\item What are your views on the future of assistive device modifications?
\end{itemize}

\subsection{Interview Guidelines for Personnel from the Accessibility Departments of the CDPF}

\subsubsection{Basic information collection}
\begin{itemize}
    \item Can you tell me your name and age?
    \item Can you tell me about your educational background?
\end{itemize}

\subsubsection{Current Structure and Services of the Disabled Persons' Federation}
\begin{itemize}
    \item What are your job responsibilities within the Disabled Persons' Federation?
\item What is the relationship between the Disabled Persons' Federation and other government departments?
\item Does the Disabled Persons' Federation have relationships with non-governmental organizations? What kind? Such as the Blind People's Association?
\item How are these relationships interconnected?
\item How is the relationship between the Disabled Persons' Federation and non-governmental organizations maintained?
\item. Is the economic support provided by the Disabled Persons' Federation to disabled people funded by government appropriations or something else?
\end{itemize}

\subsubsection{Services for Disabled People}
\begin{itemize}
    \item What services does the Disabled Persons' Federation provide to disabled individuals?
\item How does the Disabled Persons' Federation understand and confirm the needs of disabled people (communication methods with disabled individuals)? Are visits/questionnaires proactive or reactive? What are the channels used?
\item What methods are used to meet these needs?
\item What difficulties are encountered in the process of assisting disabled individuals, and how are they resolved?
\end{itemize}

\subsubsection{Identifying the Modification Needs of Disabled Individuals }
\begin{itemize}
    \item How is a tender project initiated? What is the ratio of space adaptation to assistive device modification when the Disabled Persons' Federation assists with these needs?
\item Can you provide an example of a tender and project implementation for both parts?
\item What standards must be met or approvals obtained before a tender for assistive device modification/accessibility modification can be issued? How are the modification projects and assistive device purchases selected in the tender contract?
\item Where are tenders published? How do bidders learn about tender opportunities? What is the ratio of active to passive discovery?
\item Besides tendering, what other methods does the Disabled Persons' Federation use to meet the modification needs of disabled individuals? What are the proportions?
\item In the process of meeting the modification needs of disabled individuals, where is the specific policy support manifested? What about the assistive device subsidy system?
\item What challenges and issues are encountered when the Disabled Persons' Federation undertakes work regarding assistive devices? Views on current societal issues/problems with assistive devices?
\item Are there instances where the Disabled Persons' Federation is unable to assist some disabled individuals due to procedural or systemic reasons? Can you elaborate?
\item Can you discuss the current financial subsidy system of the Disabled Persons' Federation, its tendencies, and issues?
\end{itemize}

\end{document}